# Copper-Doped Colloidal Semiconductor Quantum Wells for Luminescent Solar Concentrators

*Manoj Sharma[1,2], Kivanc Gungor[1], Aydan Yeltik[1], Murat Olutas[1,3], Burak Guzelturk[1], Yusuf Kelestemur[1], Talha Erdem[1], and Hilmi Volkan Demir [1,4,*]*

[1] Department of Electrical and Electronics Engineering, Department of Physics, and UNAM–Institute of Materials Science and Nanotechnology, Bilkent University, Ankara 06800, Turkey.

[2] Department of Nanotechnology, Sri Guru Granth Sahib World University, Punjab 140406, India.

[3] Department of Physics, Abant Izzet Baysal University, Bolu 14280, Turkey.

[4] LUMINOUS! Center of Excellence for Semiconductor Lighting and Displays, School of Electrical and Electronics Engineering, School of Physical and Mathematical Sciences, School of Materials Science and Engineering, Nanyang Technological University, Nanyang Avenue 639798, Singapore.

*Corresponding Author:

Email: volkan@stanfordalumni.org, hvdemir@ntu.edu.sg

**KEYWORDS:** Copper doping, 2D semiconductor nanoplatelets, colloidal quantum well, cation exchange, luminescent solar concentrators.

**Doping of bulk semiconductors has revealed widespread success in optoelectronic applications. In the past few decades, substantial effort has been engaged for doping at the nanoscale. Recently, doped zero-dimensional quantum dots made of semiconductor nanocrystals have been demonstrated to be promising materials for luminescent solar concentrators (LSCs) as they can be engineered for providing Stokes-shifted tunable emission in the solar spectrum. However, existing doped colloidal quantum dots suffer from moderately low quantum efficiency and intrinsically small absorption cross-section, which together fundamentally limit their effective usage. Here, we show the first account of copper doping into two-dimensional atomically flat CdSe colloidal quantum wells. Copper doping via post-synthesis partial cation exchange into 3-5 monolayers of CdSe and CdSe/CdS core-shell colloidal quantum wells (CQWs) enables Stokes-shifted, tunable dopant induced highly efficient photoluminescence emission having high quantum efficiencies (>43%), accompanied by an order of magnitude higher absorption cross-section than that of the doped quantum dots. Based on these exceptional properties, we demonstrated a flexible prototype luminescent solar concentrators using these newly synthesized doped colloidal quantum wells. These findings may open up new directions for deployment of CQWs in LSCs for advanced solar light harvesting technologies.**

Copper (Cu)-doped phosphors have been extensively investigated and used in numerous optoelectronic applications in the past century.[1] At the nanoscale, significant attention has been given to the development of Cu-doped zero- (0D) and one-dimensional (1D) colloidal semiconductor nanocrystals (NCs)[2–14] that have unique advantages including size and composition dependent tunable photoluminescence (PL) emission, almost zero self-absorption, and p-type conductivity.[15–17] These nanostructures, which generally emit in near infrared (NIR), have been extensively studied because of their possible applications sought in the fields of color conversion with phosphors, optical amplifiers, biomarkers, lasers, secure information displays and luminescent solar concentrators (LSCs).[2–16,18] For example, recently LSCs that integrate Cu-doped CdSe and CuInSe$_x$S$_{2-x}$ colloidal quantum dots (CQDs) exhibited significantly higher performance as compared to those using other colloidal NC heterostructures.[15,16,18]

CdSe CQDs have been found to be difficult for doping and, therefore, have been less explored in terms of doping as compared to other wide-bandgap II-VI NCs, e.g., ZnSe, CdS or ZnS.[5–13,19] Recently, partial and full cation exchange method has emerged as a favourable alternative to dope and produce NCs with a targeted composition or morphology that is difficult to obtain by using direct synthetic approaches.[19–22] To date, Ag$_2$Se and Cu$_2$Se NCs have been achieved through full cation exchange of the CdSe NCs.[23] Also, colloidal CQDs have been shown to be doped through post-synthesis partial cation exchange technique, where a few of the dopant cations are replaced by the host cation in the crystal lattice.[19,20] Apart from the studies on the doping techniques of Cu-doped NCs, various reports for the underlying physical mechanism of Cu-related emission have also appeared in the literature.[2–14,16,24] It is believed that Cu-related emission in doped semiconductors originates from the recombination of conduction band electron and Cu localized hole. However, there are conflicting reports for origin of this copper localized hole. These are due to different oxidation states (Cu$^{1+}$, Cu$^{2+}$) of copper reported for the doped-NC ground state.[3,5,6,24–27]

In the last few years, colloidal nanoplatelets (NPLs) have attracted great interest, which are also known as colloidal quantum wells (CQWs) due to their strong 1D confinement.[28,29] As compared to the CQDs, this new class of NCs have shown superior optical properties including narrow spontaneous emission spectra, suppressed inhomogeneous emission broadening, extremely large linear and nonlinear absorption cross-sections[30,31] and giant oscillator strengths.[32,33] In addition, assorted heterostructures of the NPLs having laterally grown crown or vertically grown shells can be synthesized with a precise control of the shell thickness and a reasonable control of the crown width.[34–36] In the NPLs, Dubertret *et al.* have recently reported full and sequential cation exchange of CdSe NPLs with Cu and then with either zinc (Zn) or lead (Pb) allowing the synthesis of NPLs having different compositions (e.g., ZnSe and PbSe, etc.).[37] However, in these fully-exchanged NPLs, PL emission has not been observed.[37] Lately, Mn(II) doping into shell region of core/shell NPLs have been shown, but dopant related emission has not been observed since energy level of the Mn stays above the bandgap of the NPLs.[38] To date, doping of Cu ions has not been demonstrated in the NPLs. Moreover, the emerging excitonic properties in the doped-NPLs and their dopant-induced PL emission has neither been achieved nor demonstrated yet.

Here, we show for the first time Cu-doped CdSe NPLs using post-synthesis partial cation exchange method. In this work, detailed optical and morphological studies have also been conducted to understand the effect of doping in varying amounts of Cu into the NPLs. The Cu-doping has been confirmed independently by different techniques including inductively coupled plasma mass spectroscopy (ICP-MS), inductively coupled plasma optical emission spectroscopy (ICP-OES), energy dispersive X-ray analysis (EDX) and X-ray photoelectron spectroscopy (XPS). With the doping of core-only NPLs having different thicknesses and core-shell NPLs, we achieved tunable dopant-induced emission covering a wide range from visible to NIR. The excitation-dependent steady-state PL spectroscopy and high-resolution XPS depth profile results have shown that Cu(I) is doped into substitutional sites in the host CdSe NPLs, which are responsible for dominant and strong dopant-related tunable PL emission with reduced self-absorption and PL quantum efficiency (QE) reaching 44%. Furthermore, the dopant-induced PL emission in the Cu(I)-doped CdSe NPLs has been corroborated by temperature-dependent time-resolved fluorescence (TRF) spectroscopy.

These Cu(I)-doped NPLs with high absorption cross-section and high PL QEs can be effectively utilized for the efficiency enhancement in the LSCs and here we demonstrate a flexible LSC prototype. We believe that these new Cu-doped colloidal CdSe quantum well nanostructures with their exceptional properties would be favorable candidates for future solar light harvesting and possibly color conversion devices.

In this work, we used a partial cation exchange method to successfully dope Cu(I) ions in CdSe NPLs having a different vertical thickness (3-5 MLs). First, we synthesized 4 ML thick undoped core CdSe NPLs having a zinc-blende crystal structure with a modified recipe.[28,29] These NPLs were used for the synthesis of Cu-doped CdSe NPLs with varying Cu concentration ranging from 0.5 to 3.6%. It has been recently reported that CdSe NPLs can be rapidly converted into $Cu_{2-x}Se$ NPLs because the d-spacing of bulk CdSe and bulk $Cu_{2-x}Se$ differ by only 5−6%, which enhances the cation exchange with Cu.[23,39] In our study, to avoid full-cation exchange, we used trioctylphosphine (TOP) as the surfactant to controllably mediate the incorporation of Cu ions into the CdSe NPLs as the dopants (see Methods section for the detailed information).

High-angle annular dark-field transmission electron microscopy (HAADF-TEM) images of the undoped and Cu-doped CdSe NPLs (for the exemplary case of 1.7 % Cu-doped) show NPLs have regular rectangular shapes with average lateral dimensions of 14.0 ± 1.5 nm × 12.9 ± 1.9 nm and vertical thickness of 1.2 nm that corresponds to 4 monolayers (ML) of CdSe (in Figure 1(a, b)). In Figure 1, it has been observed that there is no significant change in the lateral size distribution (see also Figures S1(a, b) of NPLs) and also their crystallinity (see high-resolution bright-field TEM images of undoped and 1.7 % Cu-doped NPLs in Figure S2(a, b)) after the Cu-doping process. Figure 1(c) exhibits the XRD pattern acquired from the undoped and doped NPLs and shows their characteristic peaks arising from the zinc-blende structure. In the doped sample, structural features do not change as compared to the undoped sample, which has been also observed in the selected area electron diffraction (SAED) pattern as shown in Figures S2(c, d) (See Supporting Information, section A for detailed discussions).

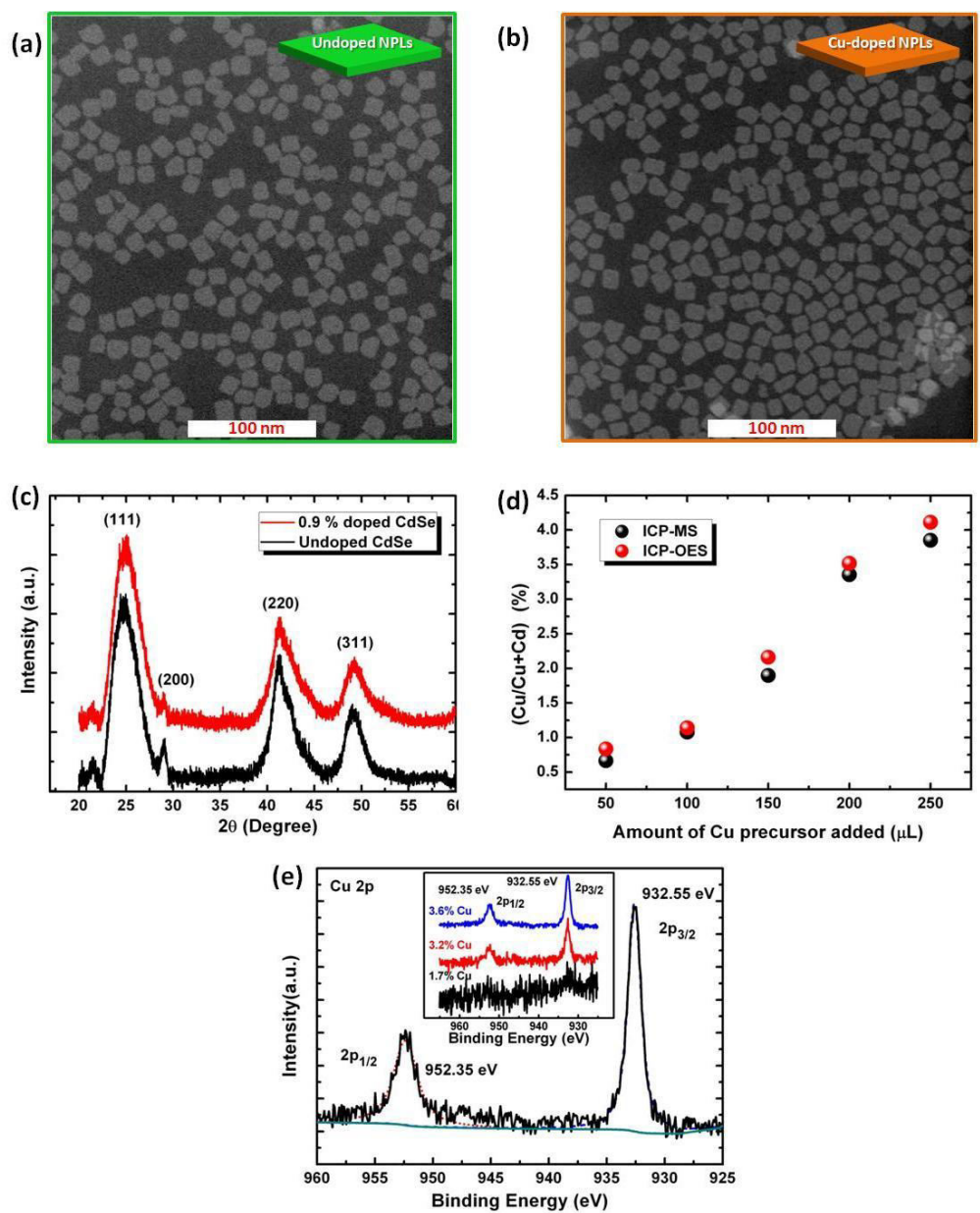

**Figure 1**. HAADF-TEM images of (a) undoped and (b) 1.7% of Cu-doped NPLs, (c) XRD pattern of undoped and Cu-doped CdSe NPLs, (d) the calculated ratio between Cu ions and total cationic ions in the NPLs with the addition of different amounts of Cu precursor during the cation exchange reaction using both ICP-MS (black solid sphere) and ICP-OES (red solid sphere) spectroscopy techniques, (e) high resolution XPS spectra of Cu 2p region (inset shows high-resolution spectra for different Cu concentrations in CdSe NPLs).

To verify the Cu doping in the NPLs, we employed ICP-OES, ICP-MS, EDX, and XPS, all consistently verifying the doping in the NPLs. EDX measurements were carried out for the undoped and (1.7 %) doped NPLs on a carbon coated nickel (Ni) grids and Cu-related peak is clearly visible in the EDX spectra of Cu-doped CdSe NPLs as presented in Figures S3. Although EDX suggests the presence of Cu in the NPLs, to determine the exact doping amounts, ICP-OES and ICP-MS were carried out. Figure 1(d) shows the change of the Cu concentration in the NPLs as a function of the Cu precursor added during the synthesis. Both ICP techniques agree very well with each other and also, there is a monotonic increase in the incorporated dopant with increasing Cu precursor (See Methods). The dopant concentration can be fine-controlled in the range of 0.5% to 3.6% by varying amount of Cu precursor solution.

XPS also shows the presence of Cu in the NPLs and furthermore provides insight to understand its possible oxidation states. For analysis, all of the peaks have been spectrally corrected according to C1s standard peak. Figures 1(e) shows the high-resolution XPS spectra of 3.6% Cu-doped NPLs for Cu 2p orbitals. The Cu 2p peaks at 932.55 and 955.35 eV given in Figure 1(e) confirm the existence of Cu(I) ions as the dopant.[6,39,40] Additional high-resolution XPS spectrum of Cu 2p for the various amount of Cu-doped samples are presented in the inset of Figure 1(e). Spin-orbital splitting of Cu(I) ions in our case is 19.8 eV, which is in good agreement with the reported value of 19.6 eV in the literature.[3,6,39] Moreover, an absence of any satellite peaks strongly suggests the absence of Cu(II) in our doped NPLs.[6,40] Therefore, these well screened peaks in our spectra indicate the presence of either Cu(I) or Cu(0) oxidation state in the CdSe NPLs. The Auger Cu LMM spectrum further offers that the valence state of copper in the NPLs should be 1+ (Figure S5(d)), which is in accordance with the recent literature of Cu-doped CdSe QDs.[23] Additional XPS analysis, survey-spectrum and high-resolution XPS for Cd and Se are shown in Figure S5 (See

Supporting Information; section B, for detailed discussion and calculation of modified Auger parameter).

Figures 2(a)-(e) depict UV-Visible absorption and steady-state PL spectroscopy results for the undoped and Cu(0.5-3.6%)-doped 4 ML CdSe NPLs. UV-Visible absorption spectra of the undoped and doped samples show that both electron-heavy hole (e-hh) and electron-light hole (e-lh) transitions of the CdSe NPLs stay unchanged after the doping, which pinpoints that the doping level studied here does not change the essential excitonic absorbance of the NPLs (in Figure 2(a)).[41] In addition, for all the doped samples, there exists a weak and broad absorption tail at lower energies below the first excitonic absorption peak (see Figure 2a). This may arise due to charge transfer (CT) absorption state, which has been previously observed in the literature for doped QDs and Type-II NC's.[2,24,34,42]

Figure 2(b) shows the PL emission spectra (dotted curves) of the undoped and Cu-doped 4 ML CdSe NPLs in solution at room temperature along with their normalized absorbance spectra (solid curves). Undoped CdSe NPL exhibits a spectrally narrow emission peak at ~514 nm having full-width at half-maximum (FWHM) of ~10 nm, which is characteristic to the CdSe NPLs having 4 ML vertical thickness.[29] For the doped-NPLs, a large Stokes-shifted broad emission at a higher wavelength region (> 650 nm) is observed in addition to the band-edge (BE) emission at ~514 nm. To understand the origin of this large Stokes-shifted emission in the doped NPLs, we performed PL excitation (PLE) measurements. Figures 2(c)-(e) show the excitation spectra of the Cu-doped NPLs with varying Cu concentration of 0.9, 1.7, and 3.2 %, respectively. The excitation spectrum of the large Stokes-shifted emission essentially resembles the absorption spectrum of the doped NPLs (see Figure 2(a)), which indicates that this large Stokes-shifted emission originates from the doped NPLs. Also, excitation spectra measured for different spectral positions of the longer-wavelength emission peak (i.e., at the peak, red- and blue-tails), which we will recall as dopant emission hereafter, do not show any discernible spectral difference (see Figure S6). This suggests that there is no inhomogeneous broadening in the doped NPL thanks to their magic-sized vertical thicknesses. As shown in Figure 3b, relative intensities between the BE and dopant emission vary as a function of Cu concentration in the doped NPLs. Dopant emission becomes monotonically stronger as the Cu concentration is increased up to 1.7%. In the case of 1.7% Cu concentration, the dopant emission has the highest overall contribution, which is an order of magnitude larger than the BE emission. For further Cu concentrations above 1.7%, we observe that BE emission starts to recover again. Apart from doping concentration, reaction time of partial cation exchange process, is also an important factor that is responsible for dominant and efficient Cu(I) related emission from NPLs (see Supporting Information, for detailed discussion, Figure S7-S9, and section C-E).

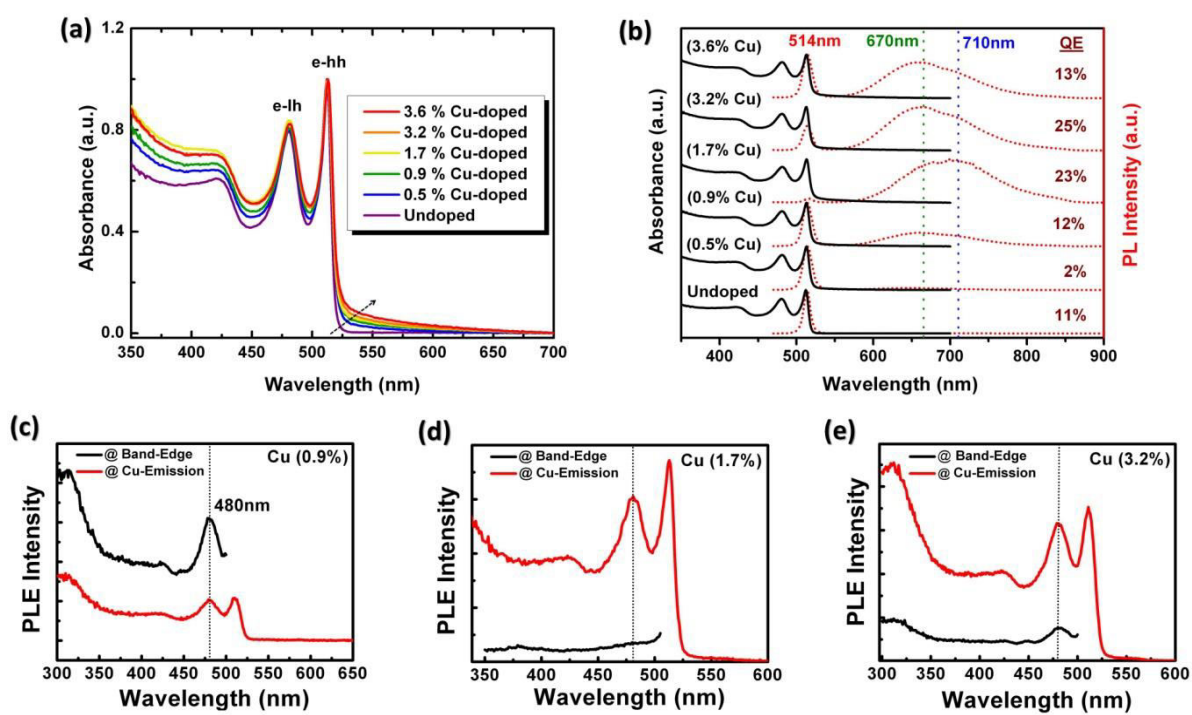

**Figure 2**. (a, b) UV-Visible absorption spectra and steady-state photoluminescence emission spectra of Cu(0-3.6%)-doped CdSe NPLs, (c, d, e) photoluminescence excitation (PLE) spectra of Cu-doped NPLs at band-edge and Cu-related emissions for different Cu-doping values ( 0.9, 1.7 and 3.2%).

We also measured the absolute quantum efficiencies (QE) using an integrating sphere based on de Mello method.[43] The PL QE of the undoped NPL is 11%. Figure 2b shows the PL QE of the doped NPLs, where PL QE varies from 2 to 25%. The highest PL QE is achieved for the doped NPL with 3.2% Cu concentration. In Table S2, we calculated the contribution of the dopant and BE emission to the total integrated QE for different amounts of doping. In the doped NPL ensembles, the contribution of the BE emission can be explained by two different hypotheses. The first is that all of the NPLs are doped and BE emission can appear from the doped NPLs, which has been shown for the Cu(II) doping in the NCs.[25] The second is that there might exist two different sub-populations within the doped ensemble: doped and undoped NCs. This second hypothesis has been recently verified by single particle studies that also reveal the existence of Cu(I) doping in CdSe NCs.[2] Also, in our doped NPLs, XPS measurements and analysis strongly suggest that the doped NPLs have 1+ oxidation state of Cu. Therefore, BE emission in the doped NPL ensembles could arise from the presence of undoped NPL fraction. Furthermore, BE emission in the doped ensemble can be used as a probe to monitor the fraction of the undoped NPL subpopulation by assuming that the PL QE of the undoped NPLs stays constant for all ensembles. To this end, in Table S2, we also estimated the undoped and doped NPL fractions in different doped NPL ensembles with varying average Cu concentration. We expect a fraction of the doped NPLs to be >80% in the ensemble. As we increase the Cu concentration, the fraction slightly increases >90%, but due to positive cooperativity principle[23], it has not been possible to achieve fully doped ensembles since already doped NCs are more prone to be further doping as compared to the undoped NCs in the ensemble. Based upon subpopulations, we also corrected the Cu doping concentrations by only considering the doped NPL subpopulation in the ensemble as shown in Table S3 (see supporting information, Section F for the detailed discussion).

To understand the fluorescence decay kinetics of the doped and undoped NPLs, time-resolved fluorescence spectroscopy using the time-correlated single photon counting system is performed (Figure 3). Figure 3a shows the fluorescence decay measured for both the dopant and BE emission. The dopant emission decay is slower than that of the BE emission. Consistent with the reported literature for undoped CdSe NPLs, the PL decay curve at the band-edge exhibits a very short lifetime, which is $\tau_{av}$ = 0.17 ± 0.01 ns, and the average lifetime for the dopant emission is 511 ± 22 ns. Indeed, the PL lifetime of the transition metal ions such as $Mn^{2+}$ and $Cu^{1+}$ is very long as stated in the previous reports for QDs.[2,6,44,45] This shows Cu(I) related emission in NPLs is similar in nature with 0D QDs.

Since the undoped NPL subpopulation should possess similar optical properties, the BE emission properties should be similar in the doped and undoped NPLs. The only difference between the undoped subpopulation in the doped ensemble and undoped NPL ensemble is that the undoped subpopulation has been exposed to TOP during the cation exchange process (see Supporting Information, Section G, H, and Figures S10 and S11). To this end, we completely evaporated excess remaining TOP at low pressure (~0.0045 mm Hg) used during TRF measurements (for details see Methods section). Then, we collected TRF decays at the peak BE emission for the temperature range of 30 – 270 K. As shown in Figure 3b, the BE emission kinetics stay the same for the undoped NPL subpopulation in the doped ensemble and the undoped NPL ensemble. If the doped NPLs could have BE emission, then we would expect to observe faster fluorescence decays in the doped NPLs, since the hole-capturing Cu state will compete with the radiative recombination of the BE excitons. Therefore, this observation supports the subpopulation hypothesis, where undoped and doped subpopulations exist in the doped NPL ensemble.

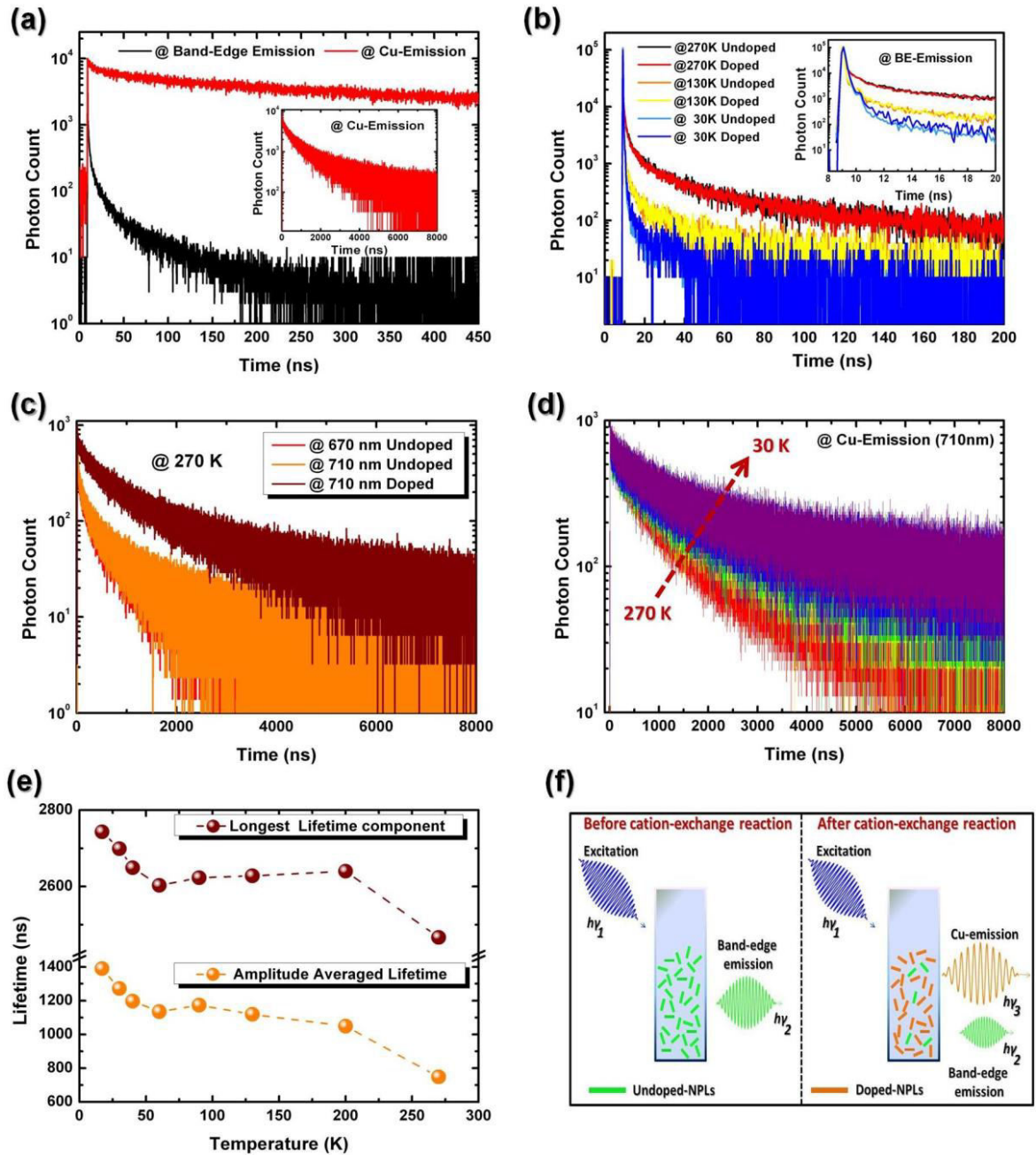

**Figure 3.** TRF decays of the undoped and Cu-doped CdSe NPLs for (a) BE and Cu-related emission at room temperature (inset shows Cu-related emission with full scale of time axis), (b) decay plots at BE emission for undoped and doped CdSe NPLs (inset shows magnified decay profile of BE related emission), (c) decay plots for trap and Cu-related emission at 270K, (d) decay profile of CdSe NPLs at different cryogenic temperatures (red@ 270K and violet @30K), (e) average and longest lifetimes variation for Cu-doped NPLs with temperature, (f) sketch showing emission from the doped and undoped subpopulations in the doped ensemble and pure undoped CdSe NPLs.

Furthermore, we compared the decay curves in the doped ensemble and pure undoped ensemble at the longer wavelength. In the pure undoped NPL ensemble, a very broad but weak defect emission has been observed. The fluorescence decay curve of this defect emission is shown in Figure 3c, where we also compared this decay to the decay of the dopant emission in the doped NPL ensemble. There is a stark contrast between the decay curves, which shows that this large Stokes-shifted emission in the doped ensemble is not due to the surface trap states but originates from the excited state between the conduction band of the CdSe NPL and the localized hole state of the Cu dopant. This observation is consistent with the previous literature on the Cu-doped nanocrystals.[2,5,6,13,24,46]

Here we also investigated temperature-dependent PL decay dynamics in the doped NPL ensemble for the exemplary case of 1.7 % Cu doping. Figure 3(d) shows the decay curves in the temperature range from 30–270 K. As the temperature was progressively decreased, we observed that the decay of the dopant emission slows down. As shown in Figure 3e, both amplitude-averaged lifetime and the longest lifetime component increase as the temperature is decreased to cryogenic temperatures. Below 60K, a monotonic increase of the fluorescence lifetime can be observed, which has been similarly reported for the Cu(I)-doped CdSe, InP and $CuInS_2$ NCs arising from the prolonged capturing probability of the photo-excited electrons in the shallow electron traps due to insufficient thermal activation energy at low temperatures.[24] A more detailed investigation of the temperature-dependent emission kinetics in the Cu-doped NPLs will be of interest to the community to understand deeper on the optical properties of these materials.

Moreover, to investigate the spectral tunability of the Cu-induced emission in the doped NPLs, we also realized Cu-doped NPLs with varying vertical thicknesses: including 3 and 5 ML, by applying the versatile partial cation exchange technique (see Methods section). Figure 4(a) exhibits the absorption and PL emission spectra of all our doped NPLs. As seen from Figure 4(a), it is conveniently possible to tune the Cu-related emission in the doped NPL ensembles from 610 to 800 nm. In addition, we also synthesized Cu-doped CdSe/CdS core/shell NPLs through the coating of 1 ML CdS shell on the undoped and doped 4 ML CdSe NPLs via colloidal atomic layer deposition (c-ALD) technique.[35] Figure 4(b) shows the absorption and PL emission of the doped core/shell NPL ensemble, where dopant emission could be red-shifted towards 750 nm, which results from the decreased confinement in the system. Thanks to the effective passivation of surface-related trap states with the CdS shell, it is also possible to increase the PL QEs further. Overall, Cu-related PL emission in the doped NPLs can effectively be tuned in the whole NIR region by doping Cu into thicker NPLs and/or having thicker core/shell nanostructures.

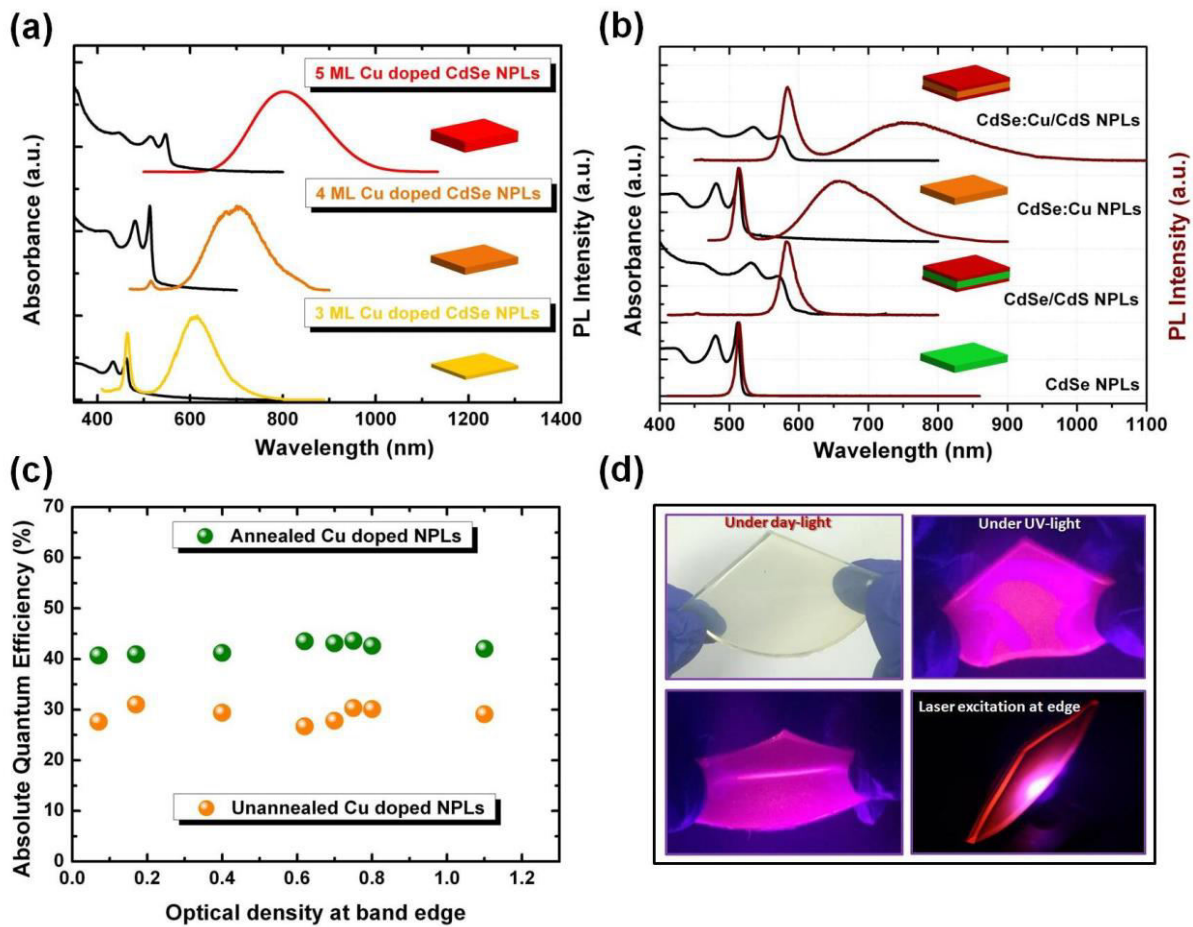

**Figure 4**. UV absorption and PL emission spectra of (a) 3 ML (yellow), 4 ML (orange) and 5 ML (red) Cu-doped CdSe NPLs, (b) undoped and Cu-doped core and CdSe/CdS core-shell NPLs, (c) absolute quantum efficiency of annealed and unannealed Cu-doped CdSe NPLs at different optical densities at BE, (d) demonstration of a prototype LSC using doped NPLs, (top left) under day light,(top right) under UV light, (lower left) showing flexible nature under UV light, (lower right) LSC emitting from edges with excitation using hand-held laser (wavelength~ 405 nm).

Among various series synthesized by the same partial cation exchange method, we achieved the highest PL QE to be 30%. More importantly, annealing of the doped NPL ensembles in the solution-state could increase their absolute QE. Cu-doped CdSe NPLs having 4 ML vertical thickness and dispersed in toluene were annealed at 120 °C for 30 min in air. We observed that PL QE increases from 30% to 44%. Time-resolved fluorescence decays before and after annealing show that the PL decay slows down, suggesting the suppression of the competing non-radiative channels (see UV, PL and TRF measurements before and after annealing, Figure S12 (a, b), and Table S4). After annealing, Cu ions might have moved to energetically more favourable positions that might overall increase the efficiency. Moreover, optimization of the experimental parameters with respect to the annealing temperature, time and environment can further lead to the increase in the absolute QE of these first ever synthesized doped NPLs.

One major advantage of the doped nanocrystals over conventional QDs is that they could considerably suppress reabsorption effect due to their very large Stokes-shifted emission.[18,47] This is a crucial property for luminescent solar concentrators (LSCs). Although the NPLs possess very large absorption cross-section, their intrinsically very small Stokes-shifted emission results in quenching of their emission in their solid-films.[48] Therefore, doped NPLs can overcome this limitation. Also, large tunability of the doped NPLs offer advantages over previously proposed systems including giant-shell CdSe/CdS QDs.[18,49] Here, we measured the PL QE of the doped NPLs with increasing optical densities. As given in Figure 4(c), even at large optical densities of 1 at the band edge, the PL QE stays constant and does not decrease due to reabsorption effect. Thus, this intriguing property makes them exciting candidates for LSCs.

Here, we also developed a prototype flexible LSC of the doped NPLs for the first time using the doped NPLs (see Figure 4d, Figure S13, and methods for LSC fabrication). For perfect LSC, apart from Stokes-shifted tunable emission in the whole NIR region, tunable absorption, high photostability, and high PL QE are the most important parameters.[15,16] In our case, the highest absolute QE obtained for Cu-doped CdSe NPLs, which is 44%, is marginally higher than the best reported Cu-doped CdSe QDs so far.[2,15,24,27] Moreover, their order of magnitude higher absorption cross-section, makes these Cu-doped NPLs very promising for LSC applications.

In conclusion, we have shown the first report for core doping of 2D CdSe colloidal quantum wells with exceptional optical properties, such as tunable emission in visible to NIR spectral region, large stokes-shift, higher PL QE's and order of magnitude larger absorption cross-section than doped QDs. Detailed XPS, temperature-dependent steady-state and time-resolved fluorescence studies strongly suggest $Cu^{1+}$ doping in CdSe CQWs. The combination of stable and dominant dopant related emission having minimum self-absorption accompanied with significantly unchanged QE's at higher optical densities confirm the novelty and advantage of deep substitutional doping of $Cu^{1+}$ in CdSe CQWs. Based on these unique properties, a flexible prototype LSC has been demonstrated. We expect doping in CQWs can be extended to different transition metal elements, which would pave the way for the synthesis of doped colloidal quantum wells and further extend their properties and applications in different optoelectronic and color conversion devices.

**Methods**

**Synthesis of 4 ML Thick CdSe Nanoplatelets-** For a typical synthesis, 340 mg of cadmium myristate, 24 mg of Se, and 30 mL of ODE were loaded into a 100 mL three-neck flask. The solution was degassed and stirred at 95°C under vacuum for an hour in order to evaporate volatile solvents and dissolve the cadmium myristate completely. Then, the heater was set to 240 °C and the vacuum was broken at 100 °C and the flask was filled with argon gas. As the temperature reaches 195 °C, the color of the solution becomes yellowish and 120 mg of cadmium acetate dihydrate has been introduced swiftly into the reaction. After the growth of CdSe NPLs at 240 °C for around 10 min, 1 mL of OA has been injected and the temperature of the solution was decreased to room temperature using a water bath. The solution was centrifuged for 5 min at 6000 rpm and the supernatant was removed into another centrifuge tube. After adding of ethanol into the supernatant solution until it becomes turbid, the solution has been centrifuged again at 6000 rpm for 10 min, and then the precipitates were dissolved and stored in toluene. The synthesis of 3 and 5 ML CdSe and CdSe/CdS core/shell NPLs has been given in the supporting information.

**Partial Cation Exchange-** All exchange reactions were carried out in an oxygen and moisture-free glove box. CdSe NPL dispersion having $2.68\times10^{-6}$ mol/L concentration has been kept under stirring at 60 °C. For a typical partial cation exchange reaction, 1ml of above CdSe NPL dispersion has been diluted to 4ml in toluene. Stock solution for copper doping was prepared by mixing 1ml of 0.4M ethanolic solution of $Cu(ac_2)_2$ with 1.5 ml TOP and then stirred for 30 min at 45 °C. It has been shown for QDs, TOP, being a soft base, binds to Cu(II) (intermediate soft acid), reducing the reactivity of dopant ions towards Se of CdSe NCs which avoids full conversion of CdSe to $Cu_2Se$ phases.[19,20] Ethanol present in the solution can help to extract replaced Cd(II) ions in the NPLs. Different amounts of stock solution were used to achieve different amount of dopant concentrations, however, the total volume of the cationic solution is made same for each reaction by adding premixed solution of TOP (3 vol%) and ethanol (2 vol%) in order to maintain a fixed concentration of TOP and ethanol for all reactions. Reactions with 50,100, 150, 200, 250μL of stock solution with 1mL of CdSe NPL dispersion ($6.7 \times10^{-7}$ mol/L) results in 0.5, 0.9, 1.7, 3.2 and 3.6 % ((Cu/Cu+Cd)%) doping of Cu(I) in 4ML CdSe NPLs. A similar method has been followed for doping of 3ML and 5ML CdSe NPLs.

Samples were stirred vigorously after each addition and the solution was allowed to equilibrate while stirring for 1-15 days. After the completion of reaction, all the samples have been precipitated with excess ethanol and washed several times using excess ethanol and dispersed in toluene. Absorption and photoluminescence spectra have been measured at different time intervals to study the effect of doping these thin NPLs.

**Fabrication of the Luminescent Solar Concentrator**

We used Platinum-cure silicone rubber (EcoFlex® 0030), which is a room temperature curable transparent and flexible rubber. Before curing the silicone rubber, we add the doped nanoplatelets that are dissolved in toluene to the silicone rubber and mix them well. Then, the mixture is poured into the mold and let for drying overnight. After the curing process, homogenously distributed, flexible and stretchable LSCs are achieved.

**Characterisations**

**TRF Spectroscopy-** Time-resolved fluorescence (TRF) spectroscopy measurements were performed by using a time-correlated single photon-counting (TCSPC) system (PicQuant FluoTime 200, PicoHarp 300). We used the picosecond pulsed laser (PicoQuant) at 375 nm for excitation, and the laser intensity was also kept low (~1 nJ/cm$^2$) so that the number of photogenerated excitons per NPL was much less than 1 (<N> << 1). In addition, The TRF spectroscopy measurements at cryogenic temperatures were carried out using a closed-cycle He cryostat that is coupled with our TRF spectroscopy system. The measurements at room temperature were conducted in solution form of NPL samples using quartz cuvette, whereas the measurements at cryogenic temperatures were carried out in solid film forms of them on Si substrates using drop-caste. On the other hand, to analyze the photoluminescence decay curves, they were fitted with multi-exponential decay functions using FluoFit software in deconvolution mode.

In our low temperature TRF measurements, pressure of the sample environment is ~0.0045 mmHg and, according to our calculations, boiling point of TOP at this pressure is ~283 K. In addition, we intentionally increased the temperature upto 330 K under vacuum and then decreased it to cryogenic temperatures. Therefore, the possible inconsistency in the emission kinetics of the NPLs due to the differences in TOP amount was eliminated by allowing the complete evaporation of TOP at low pressure conditions. In the light of this information, we could obtain the TRF decays without the interference of the effects of TOP, and at low temperatures ranging from 270 K to 17 K, which is shown in Figure 3.

**Acknowledgement**

The authors would like to thank for the financial support from EU-FP7 Nanophotonics4Energy NoE, and TUBITAK EEEAG 109E002, 109E004, 110E010, 110E217, and 112E183, and from NRF-NRFI2016-08 and A*STAR of Singapore. H.V.D. acknowledges support from ESF-EURYI and TUBA-GEBIP. M.S would like to thank TUBITAK for providing 2221 Fellowship for Visiting Scientist, BIDEB, TUBITAK, 2015. The authors wish to thank Dr Savas Delikanli for his assistance in synthesis of cores NPLs and useful discussions in the early stages of the work.

**Author contributions**

M.S and H.V.D conceived the idea and H.V.D supervised the research work. M.S carried out the colloidal synthesis, doping and optical measurements. K.G helped in LSC measurements and participated in imaging. A.Y performed time resolved fluorescence and ICP measurements. M.O synthesized (3-5ML) core NPLs and assisted in quantum efficiency measurements, PL lifetime measurements. B.G performed TEM measurements and contributed to analysis of sub-populations. Y.K assisted in NPL synthesis, XPS and XRD measurements. T.E helped in temperature stability and QE measurements. All authors contributed to discussions, analysis and writing of manuscript.

**Additional Information**

Supplementary information is available in the online version of the paper. Reprints and permissions information is available online at www.nature.com/reprints. Correspondence and requests for materials should be addressed to H.V.D.

**Competing financial interests**

The authors declare no competing financial interest.

Supporting Information

# Copper-Doped Colloidal Semiconductor Quantum Wells for Luminescent Solar Concentrators

*Manoj Sharma[1,2], Kivanc Gungor[1], Aydan Yeltik[1], Murat Olutas[1,3], , Burak Guzelturk[1], Yusuf Kelestemur[1], Talha Erdem[1], and Hilmi Volkan Demir [1,4,\*]*

[1] Department of Electrical and Electronics Engineering, Department of Physics, and UNAM–Institute of Materials Science and Nanotechnology, Bilkent University, Ankara 06800, Turkey.

[2] Department of Nanotechnology, Sri Guru Granth Sahib World University, Fatehgarh Sahib 140406, Punjab, India.

[3] Department of Physics, Abant Izzet Baysal University, Bolu 14280, Turkey.

[4] LUMINOUS! Center of Excellence for Semiconductor Lighting and Displays, School of Electrical and Electronics Engineering, School of Physical and Mathematical Sciences, School of Materials Science and Engineering, Nanyang Technological University, Nanyang Avenue 639798, Singapore.

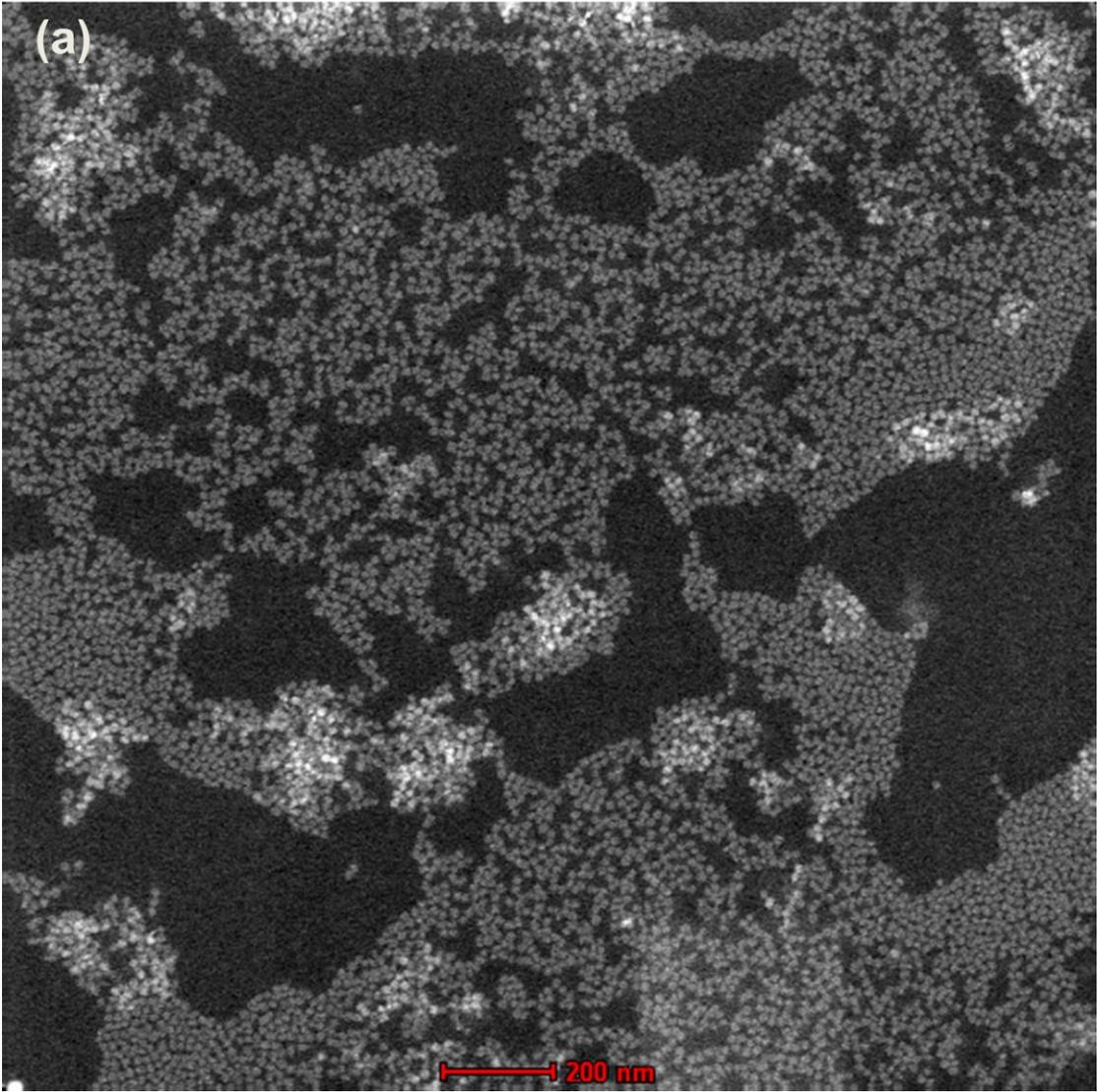

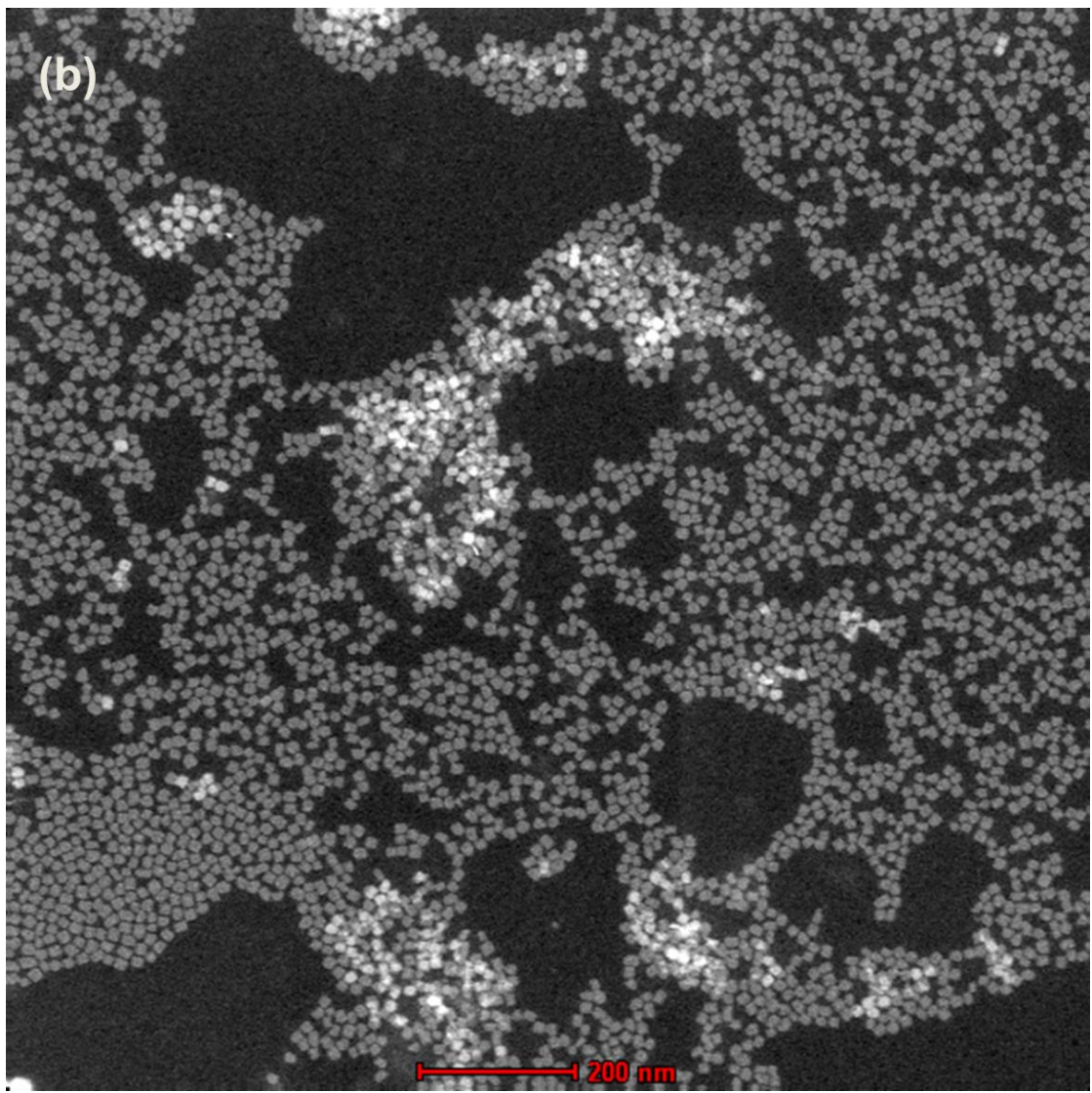

**Figure S1.** Low resolution HAADF-TEM images of (a) 0.5% and (b) 1.7% Cu doped 4 ML CdSe nanoplatelets.

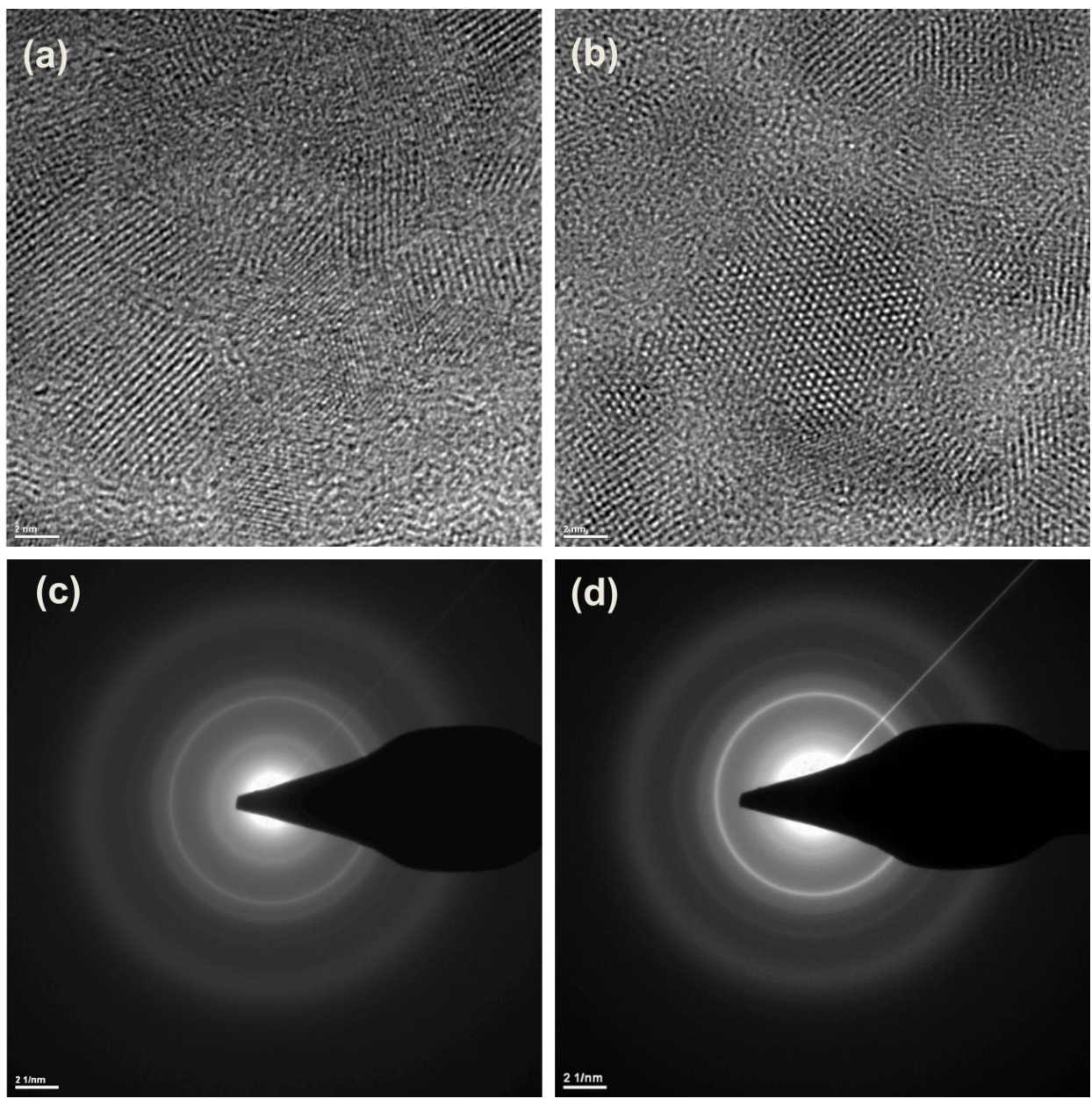

**Figure S2.** High resolution TEM (HRTEM) images and SAED patterns of (a, c) undoped and (b, d) 1.7% Cu doped 4 ML CdSe nanoplatelets. (scale bar is 2 nm for HRTEM images)

## A. Structural Analysis

TEM-based SAED measurements show that our Cu-doped CdSe NPLs are highly crystalline and they possess cubic form of the material. In particular, the Figure S2(a) and S2(b) shows lattice fringes stemming from the (111) and (220) planes of cubic CdSe.[1] The corresponding lattice spacing is 0.34nm and 0.21nm and matches well with the (111) and (220) d-spacing of bulk CdSe respectively.[1,2] In addition, the SAED measurements of the ensemble of the NPLs corroborate this conclusion as the observed rings can be indexed as (111), (220), (311), and (400) planes of the CdSe NPLs (Figure S2(d)). The d-spacing values extracted from the TEM-based SAED are 0.340 (111), 0.215 (220), 0.184 (311), and 0.151 nm (400) and they are in good agreement with the values of bulk CdSe.[3] Furthermore, undoped CdSe NPLs reveal characteristic (111) zinc-blende lattice fringes with a corresponding d-spacing of 0.34nm. Figure S2(c) presents accompanying ensemble SAED images for undoped CdSe NPLs showing the diffraction rings that can be indexed to (111), (220), (311), and (400) planes of the zinc-blende CdSe NPLs. Associated d-spacings are 0.339, 0.211, 0.180 and 0.151nm match well with those of bulk CdSe.[1] Figure 1(c) (main text) exhibits the XRD pattern acquired from the undoped and doped CdSe NPLs and reveals their characteristic zinc-blende reflections which are in agreement with ICDD card no. 19-0191.

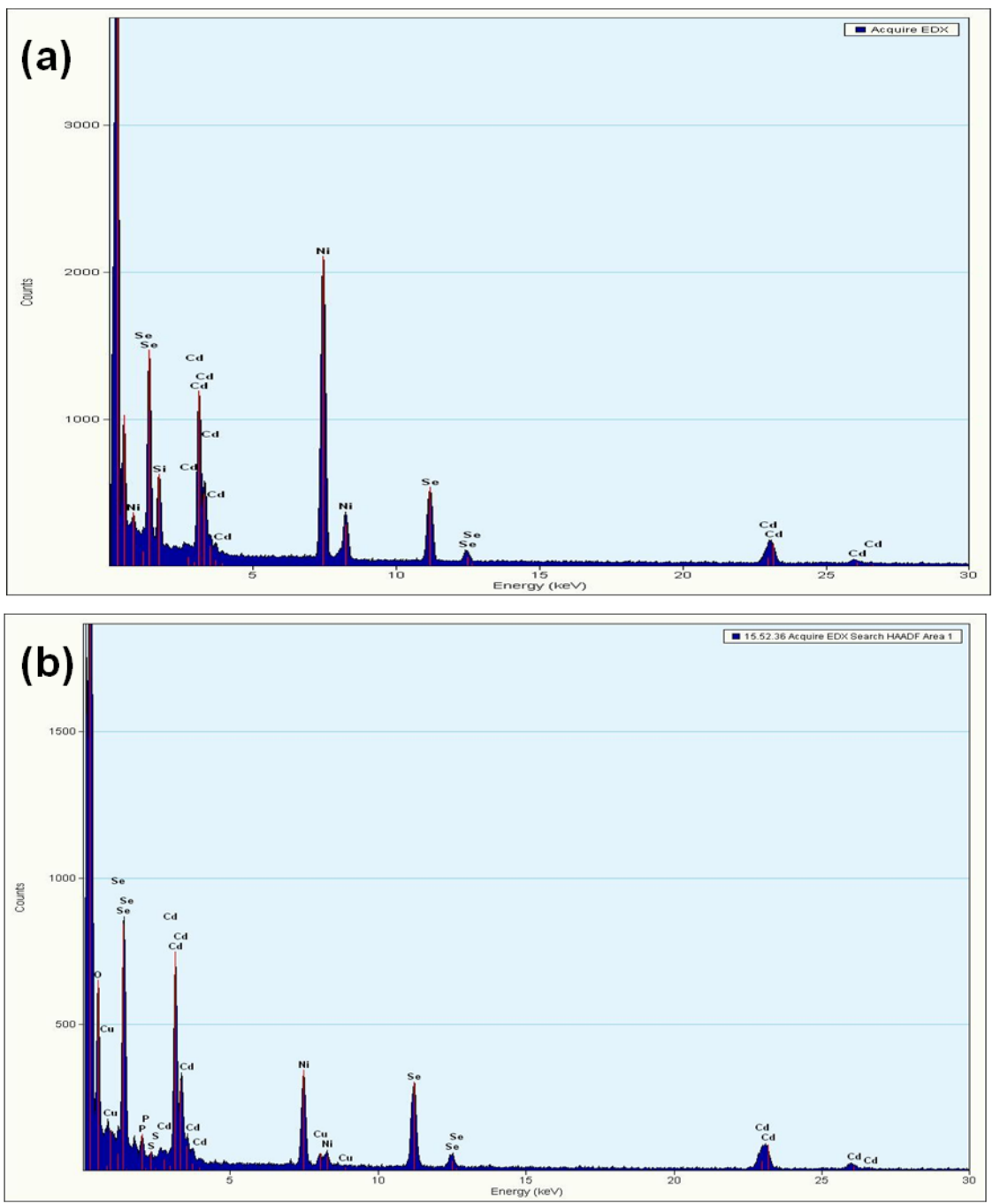

**Figure S3.** EDX spectra of (a) undoped and (b) 1.7% Cu doped 4 ML CdSe nanoplatelets.

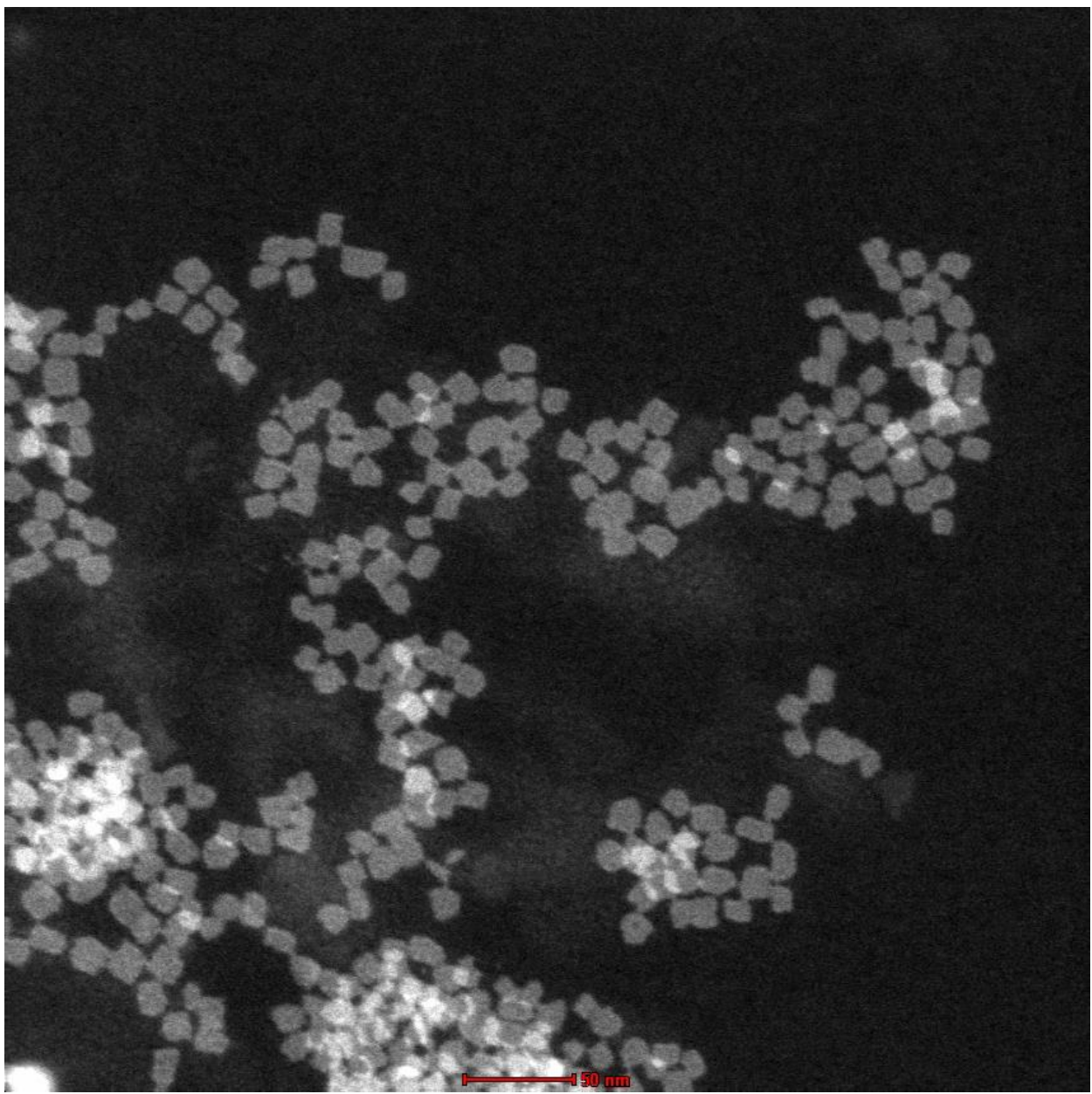

**Figure S4.** HAADF-TEM image of 3.6 % Cu- doped CdSe NPLs.

## B. XPS Analysis of Cu doped CdSe NPLs

XPS spectra of Cu doped CdSe NPLs has been shown in Figure S5(a-c). The Cu 2p peaks along with the other peaks are marked in the survey spectrum as shown in Figure S5(a). Figure S5(b) shows that the Cd 3d peak is split into $3d_{5/2}$ (405.10 eV) and $3d_{3/2}$ (411.83 eV) peaks. In Figure S5(c), the peaks at 53.79 eV and 54.61 eV correspond to the Se 3d transitions. The observed binding energies are in agreement with the reported data of CdSe NCs in the literature.[4] It was observed that the features and the positions of Cd 3d and Se 3d peaks remain almost the same regardless of the changes in the Cu concentration. Absence of satellite peaks in Cu 2p spectrum confirmed absence of Cu(II) valence state (Figure 1(e), (main text)). This shows possibility of Cu(I) or Cu(0) in our doped NPLs. In order to understand oxidation state of Cu between Cu(I) and Cu(0) we have conducted XPS based Cu Auger LMM spectra. Figure S5(d) shows the Auger Cu LMM spectrum according to which valence state of copper in the NPLs is 1+, which is in accordance with the recent literature of Cu-doped CdSe QDs.[5] Furthermore, in the absence of satellite peaks for Cu(II), the modified Auger parameter ($2p_{3/2}$, $L_3M_{45}M_{45}$) and inspection of the Auger peak-shape allow us for a more accurate assignment for difference between Cu(I) and Cu(0) which has been used effectively in literature.[6,7] Therefore, we calculated auger parameter (($2p_{3/2}$, $L_3M_{45}M_{45}$)) which matches with standard NIST database[7] and reported literature[6] showing Cu as 1+ ion in our doped NPLs.

$$\text{Auger parameter } (\alpha): = \text{Cu2P (BE)} + \text{Cu LMM (KE)}$$

$$= 932.55 + 916.64 = 1849.19 \text{ eV}$$

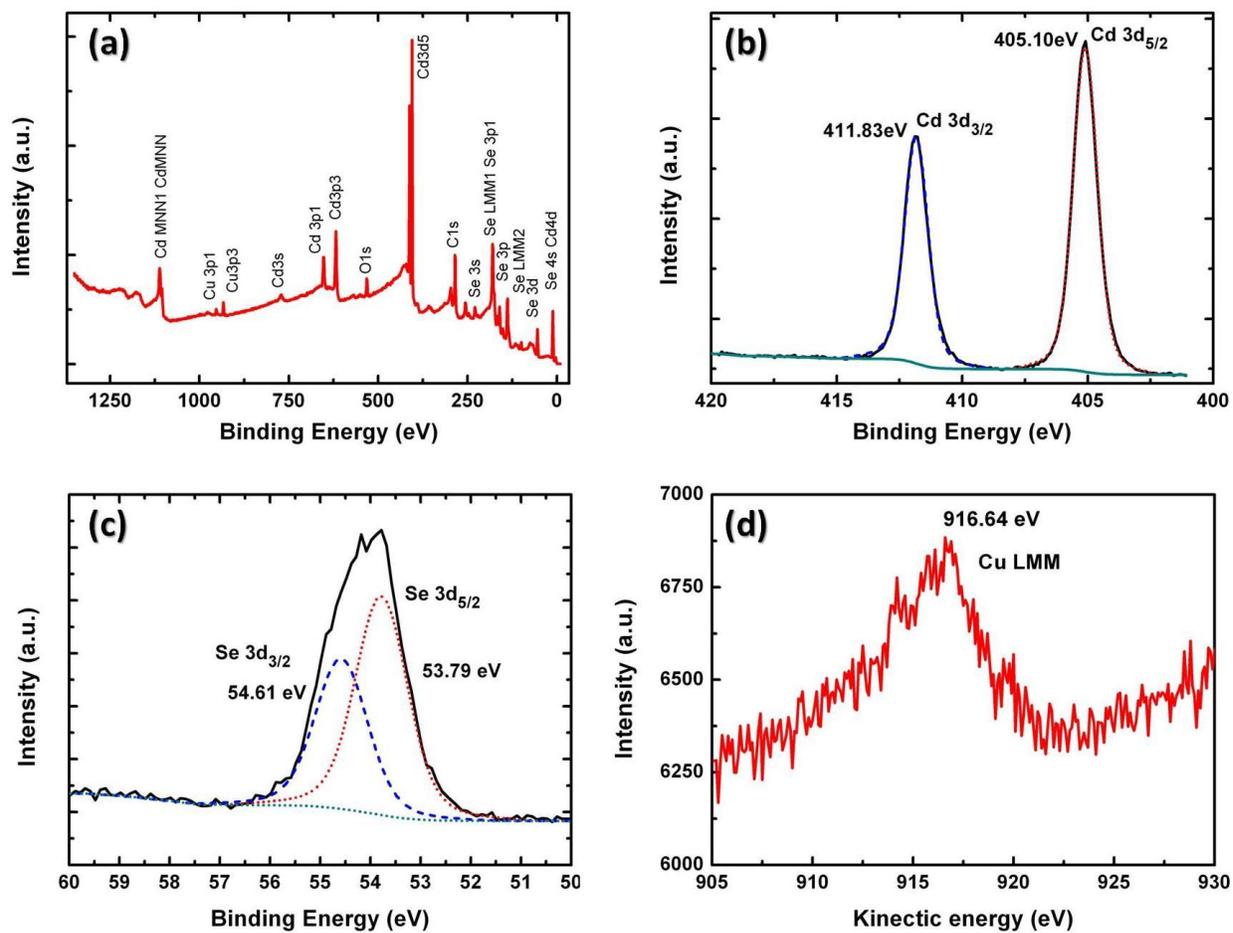

**Figure S5(a).** Survey spectrum of Cu-doped CdSe NPLs, (b, c,) high resolution XPS spectra of cadmium 3d, selenium 3d, and (d) Cu LMM Auger spectra for doped NPLs.

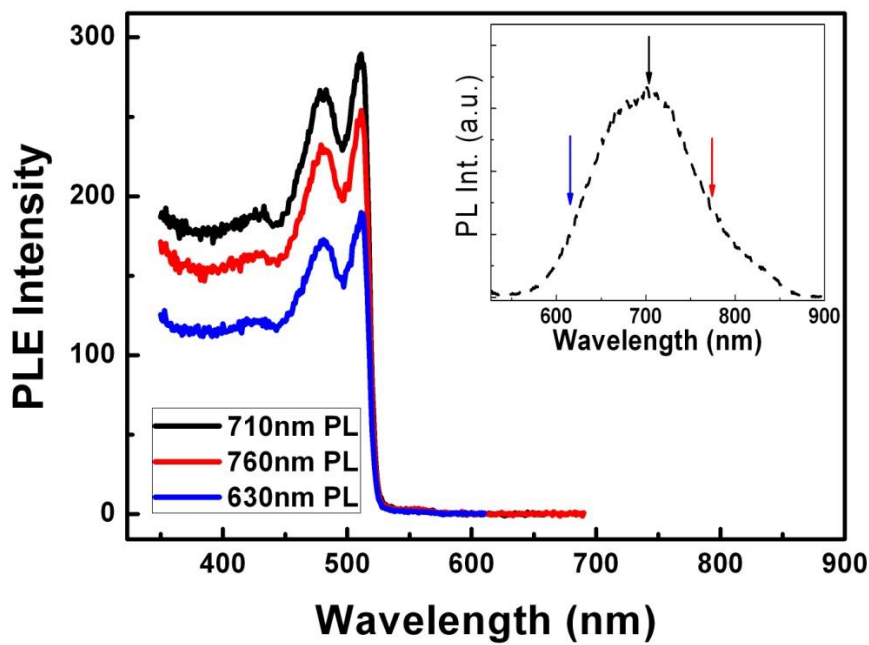

**Figure S6.** PLE spectra of 1.7 % Cu-doped CdSe NPLs at different PL emission wavelengths for Cu related broad spectrum.

### C. Reaction Time Studies for Cu-doped CdSe NPLs

In order to understand different factors for varying dopant amounts in the NPLs, we have changed the reaction time needed for the completion of cation exchange process. Figure S7 shows the UV-Visible absorption and PL emission spectra of all the samples measured after the termination of the reaction on the day-1 (Figures S7(a) and S7(b)) and day-4 (Figures S7(c) and S7(d)), respectively. For the comparison of optical studies with respect to time, we have selected three different doping concentrations of Cu precursor, which are 100, 150 and 200 μL, and we obtained 0.9, 1.7, 3.2 % Cu-doped NPL samples, respectively. Figures S8(a)-(c) show that, if the reaction is being completed after one day, PL emission spectra are similar for all the doped samples. However, the termination of the cation exchange process in fourth and ninth day creates

1.7 and 3.2 % Cu-doped NPLs, respectively. Moreover, it leads to enhanced dopant related emission along with suppression of BE emission. (Figures S8(b) and S8(c)). On the other hand, for the 0.9 % Cu-doped NPLs, there is very small increase in Cu-related emission with time. For these Cu concentrations, there is no considerable change in the PL emission spectra after the ninth day. This clearly suggests that, along with the initial copper precursor concentration, reaction time is also an important parameter for doping different amounts of Cu into the NPLs.

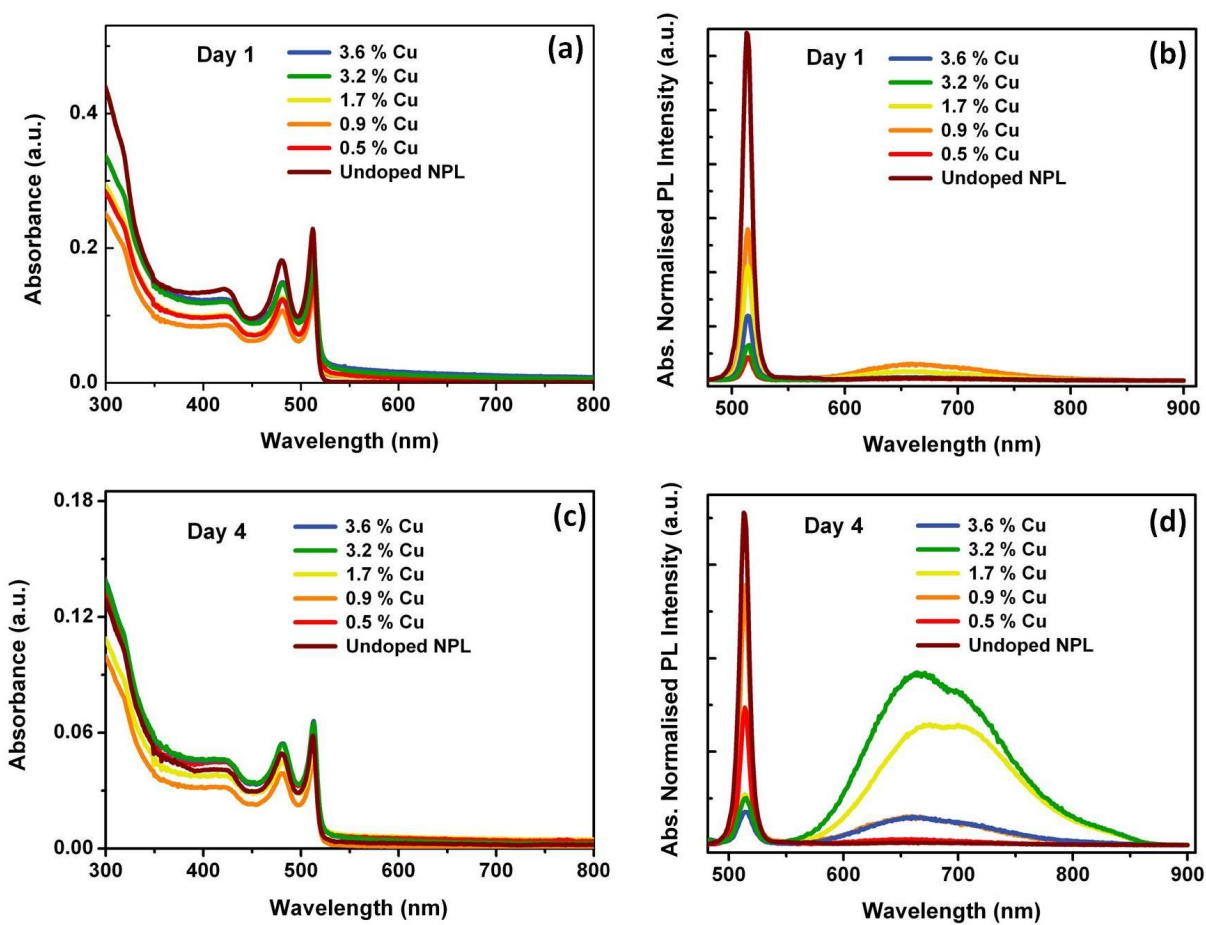

**Figure S7.** UV-Visible absorption spectra of 4 ML Cu doped samples at different times.

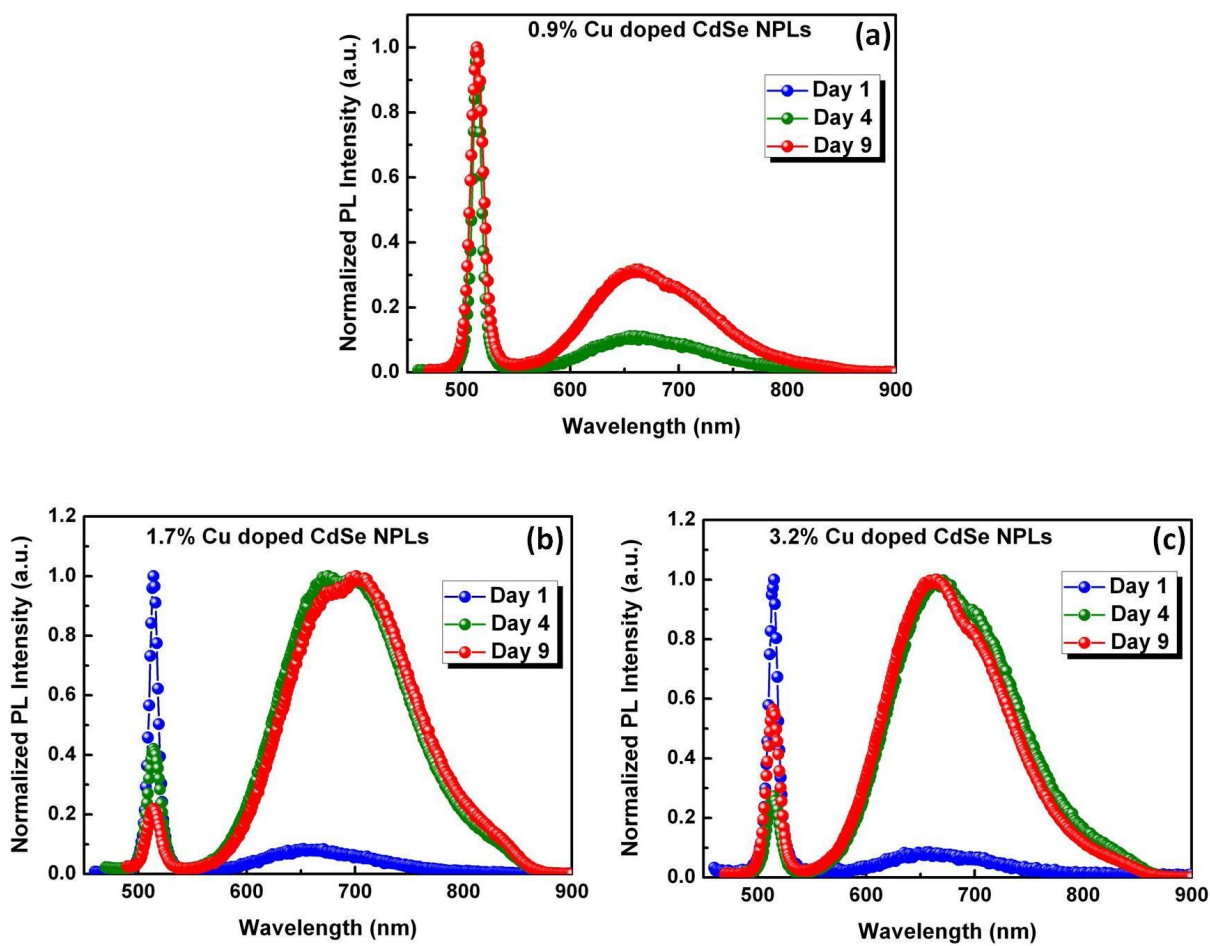

**Figure S8.** UV-Visible absorption spectra of less moderate and highly doped 4 ML Cu-doped samples at different times.

### D. Excitation Wavelength Dependent PL Emission Studies

In order to understand position of Cu ions in CdSe lattice, we have measured PL emission spectra at different excitation wavelengths (Figure S9). As confirmed by XPS analysis the presence of Cu(I) level indicate the possibility of isolated Cu(I) state within the NCs when considered as dopant. It is previously reported that Cu(I) as a dopant can form chemical bonds with CdS and CdSe constituents forming an alloyed structure with substitutional doping.[10,11] Moreover, for an isolated Cu(I) state it is believed that, the PL will solely depend on the energy gap between the conduction band (CB) edge to the Cu(I) level and the separation between Cu(I) level and the valence band (VB) edge of the NCs defines the resultant Stokes shift (Figure S9(a)).[10] Therefore, in our case of Cu doped CdSe NPLs, the energy difference between the Cu(I) level and valence band of CdSe NPLs should be about 0.68 eV (Figure S9(a)). However, our excitation dependent PL measurements exclude the existence of an only isolated Cu(I) state within the NCs. Figure S9(b) shows PL emission spectra of doped NPLs at different excitation wavelengths starting from 450 to 550 nm. Highest emission intensity is observed for 480 nm excitation. Moreover, it shows only emission intensity varies as a function of different excitation wavelengths whereas peak positions and their relative contributions remain same at all excitations. Furthermore, we have observed Cu-related emission peak at lower excitation energies than excitonic absorption i.e. for 525 and 550 nm respectively.

The appearance of dopant emission for different excitations below the band gap energy suggest the existence of additional energy levels within VB and CB of host CdSe NPLs creating a decreased energy gap for Cu-related emission. Recently for Cu doped CdS NC's, Cu emission has also been reported for excitation wavelengths lower than host band gap.[10] They have suggested this behavior with the formation of additional energy states for $Cu_2S$ between valence and conduction band of CdS. Overall, they have suggested deep substitutional doping of Cu in CdS or at highly symmetric interstitial position of Cu coordinated with sulfur.[10] Moreover, dopant induced high QE has been reported for deep substitutional doping of Ag, Cu in CdS QDs. Indeed, we have observed continuous decrease in the sub-populations of doped NPLs at higher Cu doping amounts (Table S2). Importantly, as we can see in Figures S8(b) and S8(c) with increased reaction time, Cu-related emission emerges for these moderate and high doped samples. However, it has been reported previously that surface doping in QDs can quench the luminescence therefore, to get stable dopant related emission core doping in semiconductor NCs is required.[8,12–14] This shows, in our case, for lower doped NPLs surface-doping and for moderate and higher doped NPLs core-doping is dominant. [9]

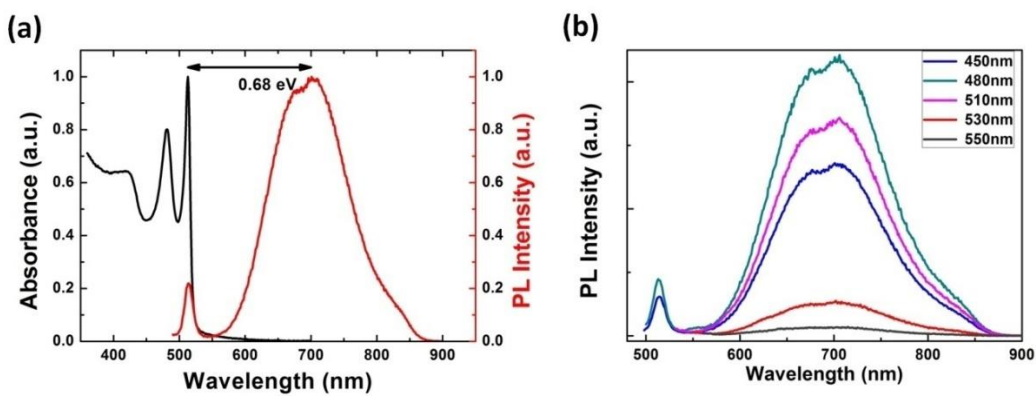

**Figure S9.** (a) UV Visible absorption and normalized PL emission spectra, (b) PL emission spectra at different excitation wavelength for Cu (1.7%) doped CdSe NPLs.

**E. XPS Depth Profile Studies**- In order to further understand possible positions of Cu as dopants inside thin NPLs, we have studied depth profile XPS for 4.6% Cu-doped NPLs. XPS depth profile shows with increase of etching cycles there is increase in copper amount from 3.45 % to 7.25 % along with constant decrease in Cadmium(Cd) amounts from 54.26 to 46.81%. Whereas there is very small variation observed in selenium concentration (from 42.28 to

45.94%) with changing Cu amounts (Table S1). It has been previously reported that undoped 4 ML NPLs are having higher Cd to Se ratio because of last terminating cadmium layer.[15,16] However, we have observed in doped NPLs there is constant decrease in Cd along with increase of Cu percentage. This implies cationic/anionic ratio decreases from 1.36 to 1.17(instead of increase for only interstitial case) with the increase of Cu-doping percentage, which in turn, suggests major percentage of Cu has been substitutionally doped along with some interstitial contributions. It is important to mention here XPS depth profile has given us ensemble elemental information from 100µm spot size (lateral) and few (2-3nm) nanometers (vertical) below. Moreover, appearance of dopant emission for different excitations below the band gap energy suggest the existence of additional energy levels within valence band(VB) and conduction band(CB) of host CdSe NPLs which results in creating a reduced energy gap for Cu-related emission. Previous reports have also suggested Cu as a substitutional dopant using X-ray absorption near edge structure (XANES) and Extended X-Ray Absorption Fine Structure (EXAFS) studies for Cu-doped CdSe and CuCdS alloyed QDs.[9–11,17] Based on our depth profile XPS studies, excitation and reaction time dependent PL emission studies along with previous literature reports, the stable and efficient Cu (I) based emission in NPLs can be accounted for deep substitutional doping. However, detailed investigations with EXAFS and XANES measurements for different Cu-doped NPLs are needed for complete understanding of dopant induced changes in internal structure of CdSe NPLs.

**Table S1.** Elemental analysis of highly doped CdSe NPLs with XPS etching

| Etch level | Cd | Se | Cu | Cd/Se | Cu/Cd | Cu+Cd | Cu+Cd/Se | Cu/Se | Cu/Cu+Cd |
|---|---|---|---|---|---|---|---|---|---|
| 9 | 46.81 | 45.94 | 7.25 | 1.0189 | 0.1548 | 54.06 | 1.1767 | 0.157 | 0.134 |
| 8 | 46.82 | 45.95 | 7.23 | 1.0189 | 0.1544 | 54.05 | 1.1762 | 0.157 | 0.133 |
| 7 | 48.07 | 44.67 | 7.26 | 1.0761 | 0.1510 | 55.33 | 1.2386 | 0.162 | 0.131 |
| 6 | 47.84 | 45.17 | 6.98 | 1.0591 | 0.1459 | 54.82 | 1.2136 | 0.154 | 0.127 |
| 5 | 48.47 | 44.66 | 6.87 | 1.0853 | 0.1417 | 55.34 | 1.2391 | 0.153 | 0.124 |
| 4 | 49.46 | 43.98 | 6.55 | 1.1246 | 0.1324 | 56.01 | 1.2735 | 0.148 | 0.116 |
| 3 | 49.92 | 44.12 | 5.95 | 1.1314 | 0.1191 | 55.87 | 1.2663 | 0.134 | 0.106 |
| 2 | 50.55 | 44.17 | 5.28 | 1.1444 | 0.1044 | 55.83 | 1.2639 | 0.119 | 0.094 |
| 1 | 53.15 | 43.05 | 3.80 | 1.2346 | 0.0715 | 56.95 | 1.3228 | 0.088 | 0.066 |
| no etching | 54.26 | 42.28 | 3.45 | 1.283 | 0.0635 | 57.71 | 1.364 | 0.081 | 0.059 |

**F. Calculation of Subpopulation in Cu(I) Doped CdSe NPLs**- For the further investigation of the optical behaviors of undoped and doped NPLs, absolute quantum efficiencies (QE) have been measured with an integrating sphere. Undoped NPLs have exhibited an absolute QE ($\eta_{PL}$) of 11%. Table S2 shows that, with doping of different amounts of Cu, absolute QE changes from 11% to the values of 2, 12, 23, 25 and 13% for the 0.5, 0.9, 1.7, 3.2 and 3.6 % Cu-doped NPLs, respectively. Herein, it is important to see the contribution of the excitonic emission to the total integrated QE with different amounts of doping. It shows with in approximation that all Cu-doped NPLs are responsible for having emission contribution in lower energies (~710 nm) then approximately 79-95 % NPLs are doped successfully in the ensemble of doped NPLs with different amounts of Cu used during cation exchange. Detailed calculated values for doped and undoped sub-populations with their relative QE are given in Table S2 using Equation 1. With this information it is clear that our ICP-MS and ICP-OES data showing Cu (0.5-3.5 %) doping in CdSe NPLs also gives ensemble doping percentages. As 6-21 % of NPLs remain undoped so it can be assumed that the actual percentage of doped NPLs in ensemble is higher than that obtained from the ICP. Based upon these measurements, if only 79-94 % NPLs are doped in ensemble, the corrected Cu doping percentages in doped sub-populations are shown in Table S3. This shows small increase in Cu doping from actual values.

Recently, it has been shown that CdSe QDs can be successfully converted into $Cu_2Se$ using cation exchange, which can also be used as a tool for studying cooperatively principle in inorganic solids.[5] This study shows that already doped QDs are more prone to be further doped and converted to new phase other than undoped sub-population in the ensemble, which is a well known phenomenon in the biological systems. Our steady state spectroscopy and QE measurement data also suggest a similar behavior (Table S2). Considering the sample with 0.5 % doping concentration level, we observed that 83.7 % of the NPLs were doped having the absolute QE of 2%. It has been shown previously that surface doping in QDs have quenched the luminescence as dopant ions would be close to surface-related defect states.[8] Therefore, the significant decrease in the absolute QE (e.g. from 11% to 2%) can be attributed to the surface doping of Cu ions in the case of lower doping concentration (e.g., 0.5%). For this reason, the 0.5

% Cu-doped sample can be excluded from analysis of sub-population calculations. Therefore, we start our analysis from 0.9% Cu-doped sample onwards. Table S1 shows with increase in doping concentration from 0.9-1.7 %, we observed a considerable increase in the population of doped NPLs from 78.8 to 94.6 % along with a significant enhancement in the absolute QE from 12% to 23%. However, with further increase in doping from 1.7 to 3.2 %, percentage population of doped NPLs is decreasing to 84.4 %. In addition to this finding, the absolute QE of this sample notably increases from 23% to 25%. Further increase in the available Cu ions, i.e. for the Cu (3.6%)-doped NPLs, leads to the decrease in the population of doped NPLs to 80.4 along with the decrease in the absolute QE to 13%. We can see from Table S2 that, with the increase in available Cu precursor amount (100-250 µL), population of doped NPLs first increases upto certain maximum limit then it starts to decrease. But elemental analysis (ICP-MS and ICP-OES) have shown constant increase in Cu doping percentage from (0.9-3.6%) with the increase in Cu precursor solution from 100 to 250 µL.

This indicates that major population of already less doped NPLs gets further doped, which has resulted in decrease in sub population of doped NPLs. As discussed above, Jain et al.[5] have shown recently, there exist a positive cooperatively for already doped NCs which are responsible for quick conversion of less doped CdSe to $Cu_2Se$ phase. For partial and full exchange with Cu(I) ions in CdSe NCs it has been shown that Cu (I) doping starts as interstitial impurities and changes to substitutional with increased dopants in core NCs. Moreover, high dopant related QE has been reported for Cu, Ag doped CdS NC's with success of doping at deep substitutional sites.[9,10] In the case of full exchange it is believed that because to the charge compensation, one cadmium vacancy is created for every two interstitial Cu (I) impurities added to lattice. Moreover, these lattice vacancies provide hopping sites for additional Cu(I) ions and leads to higher out-diffusion of Cd(II) ions from the NCs. If additional Cu(I) ions are available in reaction solution then the local negative polarization at a Cd(II) vacancy sites can electrostatically attract these positively charged Cu(I) ions. The inclusion of large interstitial impurities and formation of vacancies strains the NC lattice greatly enhancing the thermodynamic driving force for Cu(I) substitutions and structural transition to $Cu_2Se$.[5] In our case, we have carried out partial cation exchange using TOP as a soft base which has helped to perturb complete transfer. But decrease in doped population with higher Cu precursor values and higher dopant related emission in moderate and higher doped NPLs suggest already doped NPLs gets further doped which can lead to more uniform doping in NPLs from surface to inside. Moreover, these calculations suggest positive cooperatively also affects doping population in an ensemble of doped NPLs. Moreover excitation dependent PL emission spectroscopy studies discussed above (Section D) also shows signatures of deep substitutional doping which has resulted in high dopant related emission for Cu doped CdSe NPLs at moderate and higher dopant concentrations.

**Table S2.** Calculation of fraction of undoped and doped NPLs from ensemble of doped NPLs with respective QEs.

| Doping (%) | Absolute QE ($\eta_{PL}$ Ensemble) (%) | Fraction of undoped NPLs (%) ($f_1$) | QE of undoped NPLs ($\eta_1$) (%) | Fraction of doped NPLs (%) ($f_2$) | QE of doped NPLs ($\eta_2$) (%) |
|---|---|---|---|---|---|
| 0 | 11 | 100 | 10 | - | - |
| 0.5 | 02 | 16.3 | 10 | 83.7 | 04 |
| 0.9 | 12 | 21.2 | 10 | 78.8 | 13 |
| 1.7 | 23 | 05.4 | 10 | 94.6 | 24 |
| 3.2 | 25 | 15.6 | 10 | 84.4 | 28 |
| 3.6 | 13 | 19.6 | 10 | 80.4 | 13 |

$$\textbf{Quantum Efficiency (} \eta_{PL} \textbf{ Ensemble)} = (f_1 \times \eta_1 + f_2 \times \eta_2) \qquad (1)$$

**Table S3.** Corrected doping amounts estimated from calculation of sub population of doped NPLs in ensemble

| Sub population of Doped NPLs (%) | Amount of Cu precursor (µL) | Cu (%) ICP-MS Ensemble | Estimated Cu (%) for actual doped population |
|---|---|---|---|
| 83.7 | 50 | 0.5 | 2.2 |
| 78.8 | 100 | 0.9 | 1.2 |
| 93.5 | 150 | 1.7 | 1.8 |
| 91.5 | 200 | 3.2 | 3.8 |
| 84.1 | 250 | 3.6 | 4.5 |

**G. Effect of using TOP in doped NPLs and control sample (undoped NPLs)**

Figure S10 (a) shows with increase in doping amounts there has been small increase in FWHM for excitonic peak from 10-14nm for (0-3.6%) Cu doped samples. It can be attributed to use to TOP in partial cation exchange reaction. In order to understand effect of TOP on excitonic features of NPLs, UV, PL spectra for same batch with same amount of ethanol and TOP has been recorded without addition of copper. It shows broadening of excitonic features with shifting of emission wavelength to 525nm and having FWHM of 29nm (Figure S10 (b)). Also absorption features in UV Visible absorption spectra have been broadened by only use of TOP. Interestingly as shown in Figure S10 (a), with different copper-TOP concentrations has shown very less increase in BE emission broadening. It can be assigned to binding of Cu(I) ions to TOP which acts as soft base.[18–20] Since in doped samples, prior to cation exchange reactions TOP bound Cu ions facilitates copper exchange with Cadmium ions whereas in undoped samples treated with

TOP ethanol mixture only, it can directly attach to surface present cadmium sites resulting in broadening of excitonic features. However, in order to keep TOP and ethanol concentration same in all the samples we added TOP plus ethanol mixture in same ratios in all samples separately (details can be found in methods section). This excess TOP (unattached from Copper initially) can bind differently to surface cadmium in all doped samples which can lead to small increase in B.E emission (10-14nm). Since exact amount of TOP attached on CdSe surface is unknown, so for comparative studies TOP effect has to be compensated.

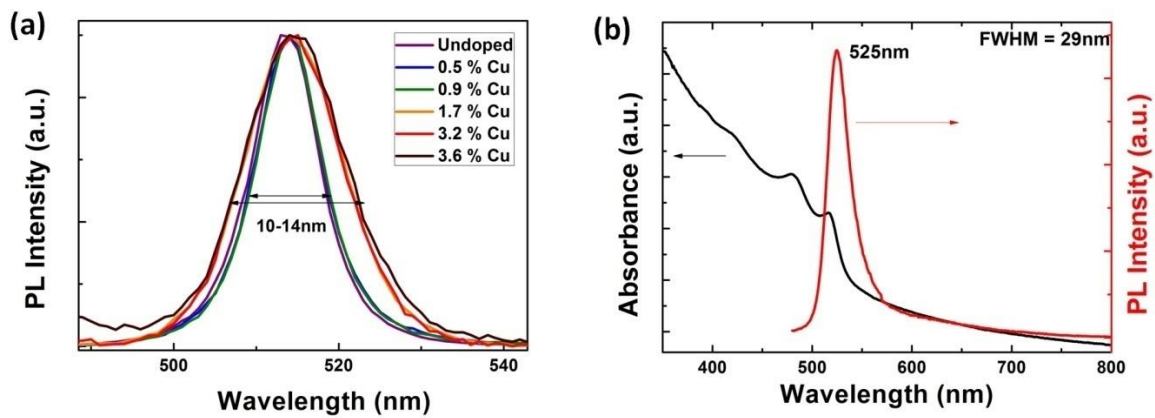

**Figure S10.** (a) Photoluminescence emission spectra of undoped CdSe and (0.5-3.6%) Cu doped NPLs for band edge emission, (b) UV Visible absorption and photoluminescence emission spectra of undoped CdSe NPLs treated with only TOP and ethanol solution.

**H. Lifetime Analysis for Understanding Effect of TOP in Doped Samples**

As aforementioned, recent work of Gamelin et al shows that excitonic emission in Cu(I)-doped CdSe QDs originates from small population of undoped QDs according to the results obtained via single particle spectroscopy.[2] It suggests that BE-related emission profile should be similar in doped and undoped NPLs. Based on these studies; small sub population of undoped NPLs (in Cu(I) doped CdSe ensemble) can be used as a control sample for the spectroscopy measurements. Therefore, comparison of excitonic features of this small sub population with actual undoped CdSe NPLs can validate above mentioned hypothesis. However, the presence of TOP in doped samples influences the excitonic features as can be seen from the Figure S10 and related discussion. Although major excitonic features are undisturbed after doping, there is small increase in the FWHM of BE-related emission from 10 to 14 nm for the doped samples using TOP (Figure S10(a)). Furthermore, it has been previously shown that increasing amount of TOP can lead to the decrease in Cu-related PL emission lifetimes.[21] For this reason, we investigated the BE- emission related decay profiles of similar TOP treated undoped and doped NPLs. It explicitly shows that there is a considerable difference in decay profiles for the TOP treated and untreated NPLs (Figure S11(a)). In addition, Figure S11(b) shows continuous increase in the Cu-related lifetime with the increasing amount of Cu-doping. Additional TOP amount (unbounded from Cu initially) is decreasing with the increase in amount of Cu-doping precursor concentrations, which can be the reason for the increase in average lifetime for the increasingly doped samples (details in methods section). Herein, the difficulty is to obtain Cu-doped NPL samples with the same amount of TOP. Furthermore, the presence of Cu attached and unattached TOP is unknown in the ensemble of NPLs during the cation exchange reactions. Therefore, since all our doped NPLs are synthesized by using Cu precursors in TOP, in order to understand the origin of BE-related emission in the doped ensemble of NPLs, effect of TOP has to be compensated for all the samples.

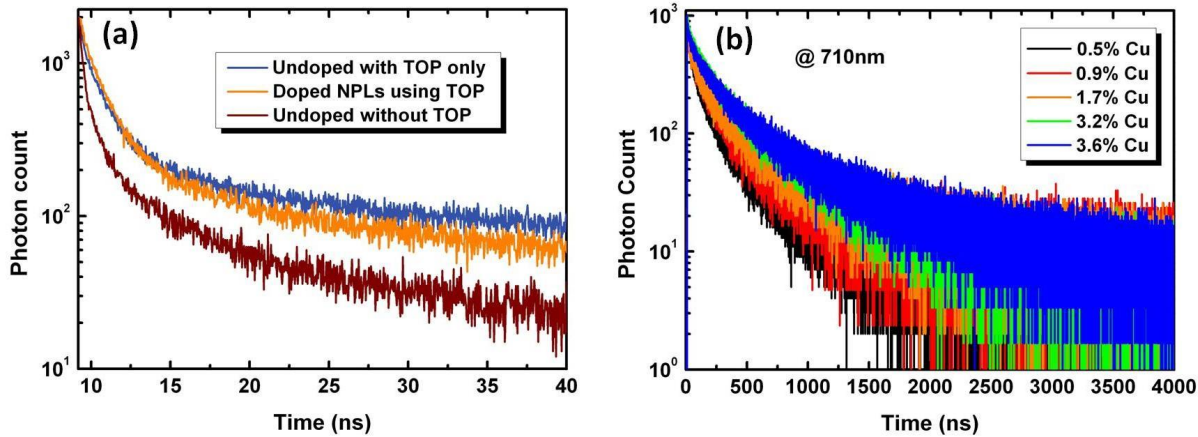

**Figure S11.** (a) Fluorescence lifetime decays curves for band edge related emission with and without TOP in doped and undoped NPLs. (b) Decay curves at Cu-related emission wavelength of the (0.5-3.6%) doped samples.

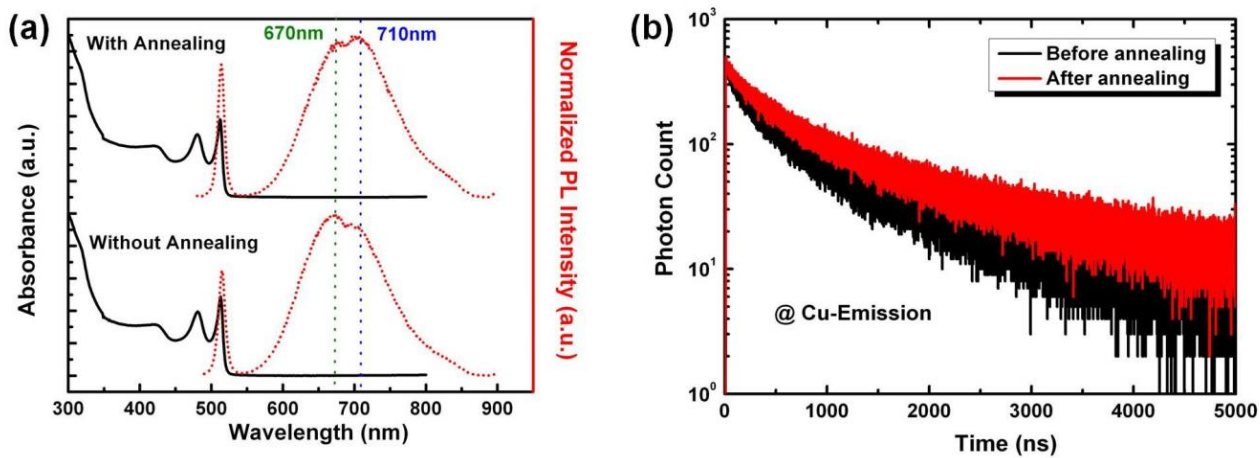

**Figure S12**. (a) UV-Visible absorption and PL emission, (b) fluorescence lifetime decays curves for Cu-related emission for Cu-doped CdSe NPLs before and after annealing.

**Table S4.** Lifetime analysis of the Cu-doped CdSe NPLs before and after annealing.

| Sample | $A_1$ | $\tau_1$ ns | $A_2$ | $\tau_2$ ns | $A_3$ | $\tau_3$ ns | $A_4$ | $\tau_4$ ns | $\tau_{av}$ ns |
|---|---|---|---|---|---|---|---|---|---|
| Before Annealing | 56.0 | 1990.6 | 162.0 | 566.4 | 163.8 | 162.7 | 81.6 | 23.3 | 500.1 |
| After Annealing | 76.3 | 2299.2 | 177.9 | 709.5 | 141.6 | 242.1 | 46.9 | 43.2 | 763.3 |

| Sample | $A_1*\tau_1$ | $A_2*\tau_2$ | $A_3*\tau_3$ | $A_4*\tau_4$ |
|---|---|---|---|---|
| | % contribution | | | |
| Before Annealing | 48.09 | 39.59 | 11.49 | 0.82 |
| After Annealing | 51.91 | 37.35 | 10.14 | 0.59 |

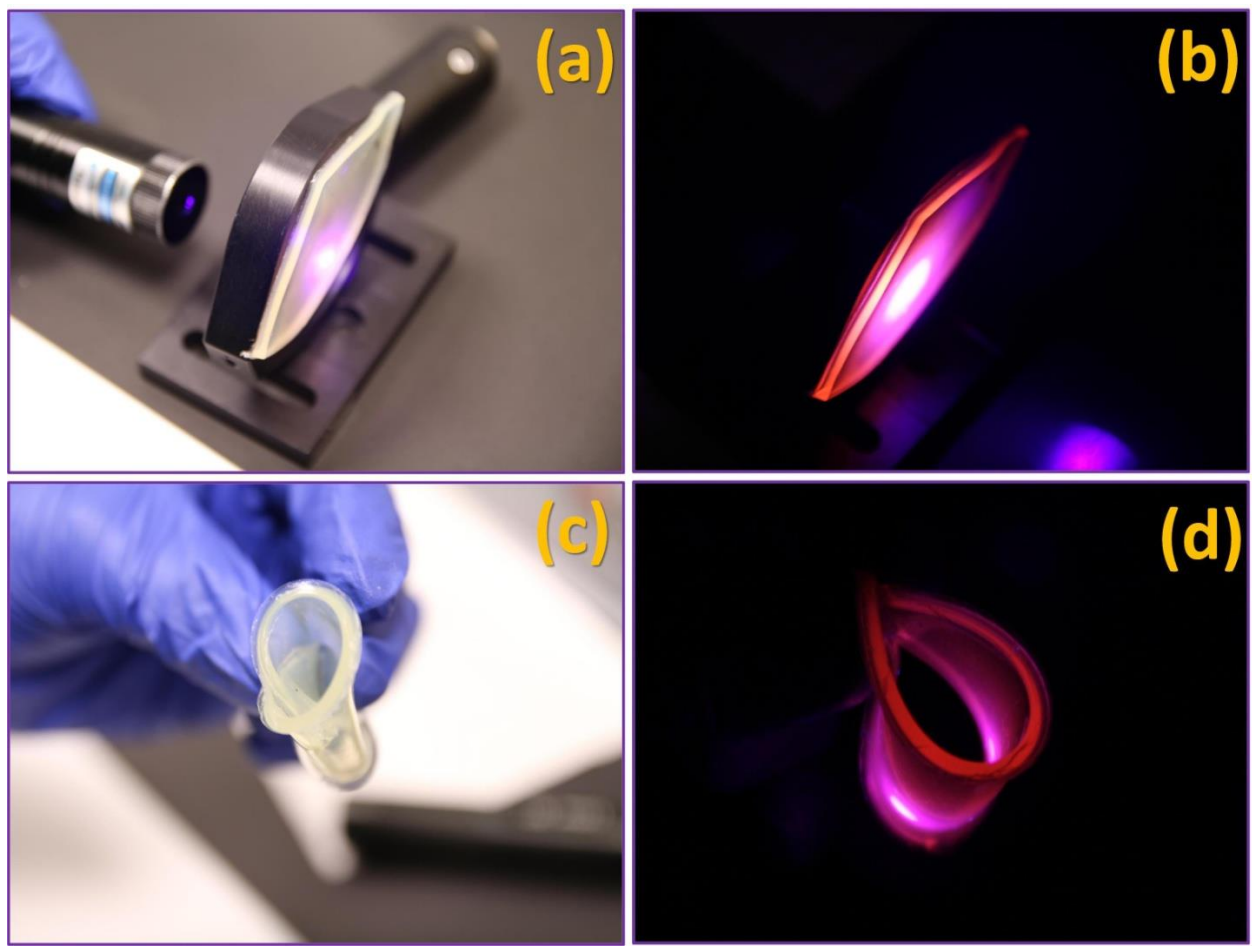

**Figure S13.** Additional images for Demonstration of Proto-type of LSC using doped NPLs, (a) under day light,(b) LSC emitting from edges with excitation using hand held laser (wavelength~ 405nm), (c) showing flexible nature under day light, (d) flexible LSC under hand held laser(wavelength~ 405nm).

**I. Additional Materials and Methods**

**Chemicals.** Cadmium nitrate tetrahydrate, sodium myristate, technical grade 1-octadecene (ODE), selenium, cadmium acetate dihydrate, copper(II) acetate, ammonium sulfide, N-methylformamide (NFA), and technical grade oleic acid (OA) were purchased from Sigma-Aldrich. Methanol, ethanol, acetone, and hexane were purchased from Merck Millipore.

**Preparation of Cadmium Myristate.** Cadmium myristate was synthesized by following the recipe given in the literature.[22] Cadmium nitrate tetrahydrate (1.23 g) was dissolved in 40 mL of methanol and 3.13 g of sodium myristate was dissolved in 250 mL of methanol under strong stirring; these solutions were then combined and stirred approximately one hour. The whitish product was centrifuged and the white precipitate part was dissolved in methanol. This washing step with methanol was performed three times for the removal of excess precursors. Subsequently, the final whitish precipitate was kept under vacuum 24 hr for drying.

**Synthesis of 5 ML Thick CdSe NPLs.** Five monolayer (5 ML) thick CdSe nanoplatelets (NPLs) are synthesized according to the recipe given in the literature.[22] 340 mg of cadmium myristate and 28 mL of ODE are loaded into a 50 mL three-neck flask. The solution is stirred and degassed at room temperature for half an hour and at 95 °C for an hour under vacuum, respectively. After heater is set to 250 °C, the vacuum is broken at 100 °C and the flask is filled with argon gas. When the temperature of the solution reaches 250 °C, a pre-prepared solution of 24 mg Se dispersed in 2 mL of ODE is swiftly injected into the hot solution. When the color of solution becomes orange, 240 mg of cadmium acetate dehydrate is rapidly introduced. Then, the solution is kept at 250 °C for 10 minutes and 1 mL of OA is injected before cooling down to room temperature using water bath. After size-selective precipitation, the 5 MLNPLs are dissolved and stored in toluene.

**Synthesis of 3 ML Thick CdSe NPLs.** Three monolayer (3 ML) thick CdSe nanoplatelets (NPLs) are synthesized according to the recipe given in the literature. [23] 217 mg of cadmium acetate dihydrate, 2 mL of pre-prepared solution of 0.15 M Se in ODE, 0.36 mL of OA, and 10 mL of ODE is loaded into a 50 mL three-neck flask. Under argon flow, when the temperature of solution reaches to 250 °C, it is kept at 250 °C for 3 min. The temperature is then rapidly cooled to room temperature. The CdSe NPLs are precipitated with the addition of acetone and dispersed in toluene.

**Calculation of CdSe NPL Concentration via Inductively Coupled Plasma Atomic Emission Spectroscopy**

In order to find out concentration of CdSe NPLs in initial and cation exchange reactions, we performed ICP-OES measurements. CdSe NPLs synthesized were cleaned several times as mentioned in methods section. After that 100μL solution of NPLs is precipitated with excess ethanol and dissolved with 2 % of 65% $HNO_3$ is added to digest the NPL for ICP measurements. ICP elemental calibration curves for Cd, Cu, and Se were created using commercial standards with known concentrations.

**Estimation of Concentration of CdSe NPLs during Partial Cation Exchange Reactions-**

Bulk density of CdSe, $\rho_{CdSe}$ = 5.81g/cm$^3$

Molecular mass of CdSe, $M_{CdSe}$ = 191.38g/mol

$N_A$ = Avogadro Number = $6.023 \times 10^{23}$ mol$^{-1}$

TEM determined lateral size, a=14nm, b=12.9nm, thickness, c=1.2 nm

Corresponding Volume, V=a×b×c= 216.72 nm$^3$

Mass of a single CdSe NPL, $m_{CdSe}$ = V×$\rho_{CdSe}$ = $1.25 \times 10^{-18}$ cm$^3$

Number of moles of CdSe per NPL, $n_{CdSe}$ = $m_{CdSe}/M_{CdSe}$ = $6.5 \times 10^{-21}$ mol

Experimental concentration of Se, [Se]= 108.9648 mg/L

NPL concentration, [NPL]$_o$ = $1.79 \times 10^{-7}$ mol/L (from ICP:OES data)

1mL stock is diluted to 15 mL for ICP

NPL stock solution concentration, [NPL]$_o$ (After correction for dilution) = $2.68 \times 10^{-6}$ mol/L

For partial cation exchange reactions we diluted this solution to 4 mL for each sample

NPL concentration in partial cation exchange reactions = $6.7 \times 10^{-7}$ mol/L